\documentclass[10pt,journal,compsoc]{IEEEtran}

\makeatletter
\def\ps@headings{%
\def\@oddhead{\mbox{}\scriptsize\rightmark \hfil \thepage}%
\def\@evenhead{\scriptsize\thepage \hfil \leftmark\mbox{}}%
\def\@oddfoot{}%
\def\@evenfoot{}}
\makeatother 
\IEEEoverridecommandlockouts
\usepackage{bbm}
\usepackage{amsfonts}
\usepackage[dvips]{graphicx}
\usepackage{times}
\usepackage{cite}
\usepackage{amsmath}
\usepackage{array}
\usepackage{amssymb}

\newcommand{\bs}{\boldsymbol}

\usepackage{stfloats}
\usepackage{graphicx}
\usepackage{footnote}
\usepackage{booktabs}
\usepackage{array}
\usepackage{algorithm}
\usepackage{subeqnarray}
\usepackage{cases}
\usepackage{threeparttable}
\usepackage{color}
\usepackage{hyperref}
\usepackage{epstopdf}
\usepackage{algpseudocode}
\usepackage{bm}
\usepackage{multirow}
\usepackage[labelformat=simple]{subcaption}
\usepackage{adjustbox}

\usepackage{enumitem}

\usepackage{xcolor,etoolbox}

\usepackage{tikz}

\let\mybibitem\bibitem
\renewcommand{\bibitem}[1]{%
\ifstrequal{#1}{8778671}{\color{black}\mybibitem{#1}}
{\ifstrequal{#1}{wang2018preempt}{\color{black}\mybibitem{#1}}
{\ifstrequal{#1}{kavitha2018controlling}{\color{black}\mybibitem{#1}}
{\color{black}\mybibitem{#1}}}}%
}

\newtheorem{theorem}{\bf Theorem}

\newtheorem{proposition}{\bf Proposition}
\newtheorem{lemma}{\bf Lemma}

\allowdisplaybreaks

\begin{document}

\title{Age of Information in Ultra-Dense IoT Systems: Performance and Mean-Field Game Analysis}

\author{Bo~Zhou,~\IEEEmembership{Member,~IEEE}
        and~Walid~Saad,~\IEEEmembership{Fellow,~IEEE}
\IEEEcompsocitemizethanks{\IEEEcompsocthanksitem B.~Zhou is with the College of Electronic and Information Engineering, Nanjing University of Aeronautics and Astronautics, Nanjing, China.
W.~Saad is with Wireless@VT, Bradley Department of Electrical and Computer Engineering, Virginia Tech, Blacksburg, VA 24061, USA.
Part of this work was done when the first author was with Wireless@VT, Bradley Department of Electrical and Computer Engineering, Virginia Tech, USA.
\protect\\
Email: b.zhou@nuaa.edu.cn, walids@vt.edu.}
\thanks{This work was supported in part by the Natural Science Foundation of China under Grant 62201255 and in part by the Office of Naval Research (ONR) under MURI Grant N00014-19-1-2621.}}
\IEEEtitleabstractindextext{%
\begin{abstract}
In this paper, a dense Internet of Things (IoT) monitoring system is considered in which a large number of IoT devices contend for channel access so as to transmit timely status updates to the corresponding receivers using a carrier sense multiple access (CSMA) scheme. 
 Under two packet management schemes with and without preemption in service, the closed-form expressions of the average age of information (AoI) and the average peak AoI of each device is characterized.
It is shown that the scheme with preemption in service always leads to a smaller average AoI and a smaller average peak AoI, compared to the scheme without preemption in service.
Then, a distributed noncooperative medium access control game is formulated in which each device optimizes its waiting rate so as to minimize its average AoI or average peak AoI under an average energy cost constraint on channel sensing and packet transmitting.
 To overcome the challenges of solving this game for an ultra-dense IoT, a mean-field game (MFG) approach  is proposed to study the asymptotic performance of each device for the system in the large population regime. The accuracy of the MFG is analyzed, and the existence, uniqueness, and convergence  of the mean-field equilibrium (MFE) are investigated.
Simulation results  show that the proposed MFG is accurate even for a small number of devices; and the proposed CSMA-type scheme under the MFG analysis outperforms three baseline schemes with fixed and dynamic waiting rates.
Moreover, it is observed that the average AoI and the average peak AoI under the MFE do not necessarily decrease with the arrival rate.
\end{abstract}


\begin{IEEEkeywords}
Internet of things, game theory, age of information, mean field, optimization.
\end{IEEEkeywords}
}

\maketitle

\IEEEdisplaynontitleabstractindextext
\IEEEpeerreviewmaketitle

\IEEEraisesectionheading{\section{Introduction}\label{sec:introduction}}
\IEEEPARstart{T}{he} proliferation of the Internet of Things (IoT) applications\cite{sisinni2018industrial,8869705,9060999,zanella2014internet}, such as, environmental monitoring, autonomous driving,  and smart surveillance, has significantly boosted the need for real-time status information updates of various real-world physical processes monitored by a large number of IoT devices.
To measure the freshness of the status information in such time-critical IoT applications, the concept of \emph{age of information} (AoI) has recently become  a fundamental performance metric in communications systems \cite{Pappas2022,6195689,
8469047,7415972,8323423,8945230,8000687,8695040,8822722,abd2020aoi,8778671,8938128,najm2016age}.
However, next-generation IoT systems will be massive in scale and will encompass a very large number of devices\cite{9060999,zanella2014internet}. In such ultra-dense IoT systems, it is often impractical to perform coordinated channel access with a centralized unit (e.g., an access point or a base station), as it requires time synchronization among all the devices and a significant amount of signaling overhead\cite{zucchetto2017uncoordinated,9060999}.
Therefore,  for ultra dense IoT systems, it is of great importance to study the AoI performance under  uncoordinated channel access and investigate how to optimize the channel access strategies in a distributed way so as to minimize the AoI.
In order to do so, several challenges must be overcome such as the characterization of the complex temporal evolution of the AoI under uncoordinated access and the strong coupling (in terms of access resources) among a large number of IoT devices.

\subsection{Existing Works}
Recently, the problem of analyzing and minimizing the AoI under uncoordinated channel access has attracted increasing attention\cite{8006544,talak2018distributed,Chen2019,Chen2020,yang2021understanding,jiang2018can,8901143,gopal2020non,bedewy2019optimizing,9007478}.
Generally, these works can be classified into two broad groups based on the type of the uncoordinated access schemes.
The first group in \cite{8006544,talak2018distributed,Chen2019,Chen2020,yang2021understanding} considers slotted ALOHA-like random access schemes, under which each device attempts to transmit its status update to the receiver in each slot with a certain probability.
Specifically, the works in \cite{8006544} and \cite{talak2018distributed} study the optimal attempt probabilities that can minimize the  average AoI in a multiaccess channel and a general wireless network with pairwise interference constraints, respectively.
In \cite{Chen2019}, the authors  propose an age-based decentralized dynamic transmission policy to minimize the total average AoI.
In \cite{Chen2020}, the authors  consider an age-based distributed  randomized  transmission policy and derive an approximate closed-form expression of the average AoI.
The work in \cite{yang2021understanding} studies the peak and the average AoI performance for Poisson bipolar modeled  large-scale wireless networks.
Note that, in a real-world dense IoT, ALOHA-like random access schemes would result in significant collisions, effectively rendering communication unreliable\cite{zucchetto2017uncoordinated,9060999}.
In this regard, to reduce the collisions suffered by ALOHA schemes, the second group in \cite{jiang2018can,8901143,bedewy2019optimizing,gopal2020non,9007478,zhou2020performance} considers the carrier sense multiple access (CSMA) type schemes, by enabling carrier sensing capabilities at the devices.
In particular, the work in \cite{jiang2018can} proposes an index-prioritized distributed random access policy to minimize the total average AoI in a status update system.
The work in \cite{8901143} analyzes the worst case of the average AoI and the average peak AoI from the view of a particular device with random status packets arrivals, for a system in which all other devices always have status packets to send.
The work in \cite{gopal2020non}  considers a status updating system in which all devices always have fresh status packets to send, and studies
 a noncooperative  game in which multiple devices contend for the spectrum using a CSMA-type scheme so as to minimize their own instantaneous AoI in one transmission.
In \cite{bedewy2019optimizing}, the authors propose an efficient sleep-wake strategy to  minimize the total average peak AoI for a wireless network in which each device uses carrier sensing and generates status packets at-will.
In \cite{9007478}, the authors investigate the problem of optimizing the backoff times of the devices so as to minimize the total average AoI.
In \cite{zhou2020performance}, we studied the average AoI in a ultra-dense IoT with noisy channels and analyzed the effects of  system parameters on the average AoI.

There are two limitations in prior works on the analysis and optimization of the AoI for CSMA-type access schemes \cite{jiang2018can,9007478,8901143,bedewy2019optimizing,gopal2020non,zhou2020performance}. First, \emph{there is a lack of concrete analytical characterization} of the average AoI and the average peak AoI for ultra-dense IoTs with random status packets arrivals, under CSMA.
Note that the works in \cite{8901143} and \cite{9007478} have analyzed the AoI performance for systems with random status updates arrivals.
However, the work in \cite{8901143}  studies the worst case of the average AoI and the average peak AoI by assuming that only the device of interest has random status updates arrivals and all other devices always have fresh status updates to send, and the work in \cite{9007478} focuses on the upper bound of the average AoI by using the technique of ``fake updates'' \cite{8469047}.
Second, \emph{no prior work provided a framework for optimizing the average AoI and the average peak AoI for each device in an ultra-dense IoT system} with random status updates arrivals, by taking into account the selfish behavior of each device.
Note that the works in \cite{jiang2018can,bedewy2019optimizing}, and \cite{9007478} consider the minimization of the AoI from the perspective of the entire system, i.e., the total average AoI or average peak AoI of all devices, and, thus, do not consider the distributed, selfish behavior of each device. Meanwhile, the work in \cite{gopal2020non} focuses only on the minimization of the instantaneous AoI of each user in one transmission by assuming that each device always has a fresh status update to send.
The work in \cite{zhou2020performance} focuses only on the analysis of the average AoI  and does not consider the minimization of the average AoI and the average peak AoI.
Note that the performance of large-scale wireless networks has also been extensively studied in the literature under CSMA \cite{5340575,10.1145/2825236.2825241,10.1145/3323679.3326498,6239591}. However, these works mainly focus on analyzing the conventional performance metrics such as throughput and delay, and did not consider the AoI performance, which is known to be a fundamentally different performance metric compared with conventional throughput or delay \cite{6195689}.

\subsection{Contributions}
The main contribution of this paper is, thus, \textcolor{black}{an accurate approximate} characterization of the average AoI and the average peak AoI for an ultra-dense IoT system with random status packets arrivals under a CSMA-type access scheme, coupled with a game-theoretic framework that enables each device to optimize its access strategy so as to minimize its own average AoI and average peak AoI.
In particular, we consider a CSMA-type distributed random access scheme for an ultra-dense IoT monitoring system, under which multiple IoT devices contend for channel access and transmit their status updates to the corresponding receivers.
Instead of transmitting a newly arrived status update immediately, each device must first sense a wireless channel to determine whether it is idle or busy.
If a channel is sensed busy, then the device needs to wait (or backoff) for an exponentially distributed time duration.
We analyze the AoI performance of each device for two packet management schemes with and without preemption in service, and we derive the closed-form expressions of the average AoI and the average peak AoI for both schemes.
We show that the scheme with preemption in service always achieves better AoI performance (in terms of both the average AoI and the average peak AoI) compared to the scheme without preemption in service.
Then, for both packet management schemes, we formulate a distributed noncooperative  game \cite{han2019game} using which, each device determines its optimal waiting (backoff) rate to minimize its average AoI or its average peak AoI  under an average energy cost constraint on channel sensing and packet transmitting.
The game is then shown to be intractable for a large number of devices.
To overcome this challenge, we propose a mean-field game (MFG) framework to study the asymptotic performance of each device for the considered IoT monitoring system in the large population regime, by using a mean field approximation\cite{10.1145/2825236.2825241,10.1145/3084454,10.1145/2964791.2901463}.
We analyze the accuracy of the proposed MFG and present a comprehensive analysis of the existence, uniqueness, and convergence of the mean-field equilibrium (MFE).
Extensive simulation results validate our analytical results and show the effectiveness of the proposed CSMA-type scheme under the MFG over three baseline schemes. Particularly, we observe that the proposed MFG is very accurate even for a small number of devices. Moreover, we also show that the average AoI and the average peak AoI under the MFE for the two packet management schemes do not necessarily decrease with the arrival rate.
In summary, the obtained analytical and simulation results provide novel and holistic insights on the analysis and optimization of the AoI performance in practical ultra-dense IoT systems.

The rest of this paper is organized as follows.
Section~\ref{sec:systemmodel} presents the system model and the CSMA-type random access scheme.
In Section~\ref{sec:aoi_analysis}, we analyze the average AoI and the average peak AoI  for two  packet management schemes.
In Section~\ref{sec:mfg}, we propose a MFG framework and analyze the properties of the MFE.
Section~\ref{sec:simulations} presents and analyzes numerical results.
Finally, conclusions are drawn in Section~\ref{sec:conclusion}.

\section{System Model}\label{sec:systemmodel}

\begin{figure}[!t]
\begin{minipage}[h]{01\linewidth}
\centering
       \includegraphics[scale=0.75]{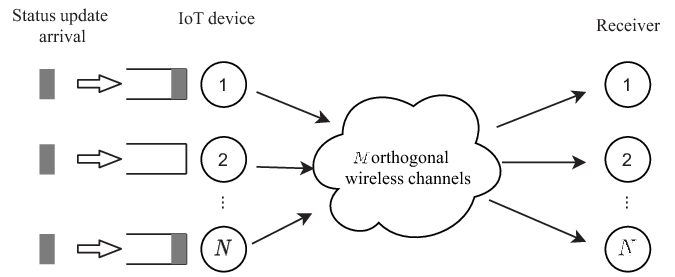}
\subcaption{}\label{fig:system_model}
\end{minipage}

\begin{minipage}[h]{1\linewidth}
\centering
        \includegraphics[scale=1]{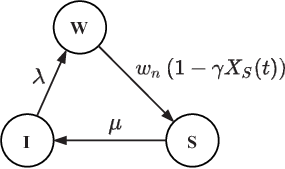}
\subcaption{}\label{fig:ctmc}
\end{minipage}
\caption{(a) Illustration of a real-time IoT monitoring system with $N$ IoT devices and $M$ orthogonal wireless channels. (b) Illustration of the state transition diagram of the CTMC $\{Q_n(t),t\geq 0\}$ of IoT device $n$.}\label{fig:system}
\end{figure}

Consider a real-time ultra-dense IoT monitoring system consisting of  a set $\mathcal{N}$ of $N$ identical IoT devices and $M$ identical orthogonal  wireless channels, as illustrated in Fig.~\ref{fig:system_model}. Let $\gamma=N/M$.
Each  device monitors the associated underlying time-varying physical process and transmits the real-time status information to its corresponding receiver.
The status information updates of each underlying process arrive randomly at each  device and may be queued at the device before being transmitted to the receiver or discarded according to the packet management scheme.
We consider that each device can  occupy at most one channel during the transmission of one status update and each channel can only be occupied by at most one device at each time instance to avoid collisions.
As is commonly done in the literature of status update systems, e.g., \cite{7415972,8323423,8469047}, we assume that the status arrival process at each IoT device follows a Poisson process of rate $\lambda$ and the transmission time of each status update is an exponentially distributed random variable with mean $1/\mu$.

We consider a CSMA-type distributed random access scheme, in which, each IoT device listens the channels before transmitting a status update.
As commonly done in the literature of CSMA (e.g., \cite{10.1145/2825236.2825241,9007478,10.1145/3323679.3326498,5340575,zhou2020performance}), we adopt idealized CSMA assumptions, i.e., channel sensing is instantaneous and there are no hidden nodes. Thus,  the probability that multiple devices transmit their packets over any given channel concurrently is zero.
\textcolor{black}{Note that sensing channels and transmitting status packets incur certain energy cost for the devices. For each device, let  $C_s$ and $C_t$ be the channel sensing cost per unit time and the transmission cost per unit time, respectively. To model the limited energy of  each device, we impose a constraint on the average energy cost per unit time denoted by $\hat{C}$.}

Next, we describe the state dynamics of each device in our system under CSMA. For each IoT device, an arriving status update may find the device in three different states: i) Idle (I), ii) Waiting (W), and iii) Service (S).
\begin{itemize}
    \item
    If an arriving status update finds an IoT device in state (I), then prior to transmitting it immediately, the device first senses one channel to check whether it is busy or idle.
If this channel is sensed idle, then the device waits (or backs off) for a random duration of time that is exponentially distributed with rate $w_n$ and then starts its transmission.
During the backoff period, the device keeps sensing the channel to identify any conflicting transmissions. If any such transmission is sensed, then the device suspends its backoff timer and waits for the channel to be idle to resume it.
Note that, this scheme of ``waiting before transmitting'' shares similar merits with those in \cite{8945230} and \cite{8000687}.
 \item If an arriving status update finds an IoT device in state (W), then this fresh status update will replace the older one that is waiting at the device, as the receiver will not benefit from acquiring an outdated status update.
 \item
 If an arriving status update finds an IoT device in state (S),  as in \cite{8469047} and \cite{7415972}, we consider two packet management schemes depending on whether the status update currently in service can be preempted. The first one is a  \emph{scheme with preemption in service} in which the status update currently in service will be preempted by the newly arrived one and then be discarded. The second one is a \emph{scheme without preemption in service} in which the  newly arrived status update will be discarded.
\end{itemize}


Let $Q_n(t)\in\mathcal{Q}\triangleq\{I,W,S\}$ be the state of device $n$ at time $t$.
Then, let $X_q(t)$ be the fraction of devices in state $q\in\mathcal{Q}$ at time $t$, given by:
\begin{align}
X_q(t) \triangleq \frac{1}{N}\sum_{n=1}^N \bs{1}\left(Q_n(t)=q\right),\label{eqn:population process}
\end{align}
where $\bs{1}(\cdot)$ is the indicator function. Define $\bs{X}(t)\triangleq (X_q(t))_{q\in\mathcal{Q}}$.
The state dynamics $\{Q_n(t),t\geq 0\}$ of each device $n$ will be a finite state continuous-time Markov chain (CTMC).
When the device is in state (I), if there is a new arriving status update, then it will transit to state (W) with rate $\lambda$.
When the device is in state (W), the probability that any channel is sensed busy is $\frac{1}{M}NX_S(t) = \gamma X_S(t)$. Thus, the transition rate from state (W) to state (S) is   $w_n(1-\gamma X_S(t))$.\textcolor{black}{\footnote{\textcolor{black}{Note that $X_S(t)$ is the fraction of IoT devices occupying the channels. As there will be at most $M$ IoT devices occupying the channels, we have $X_S(t)\leq M/N$, which implies $\gamma X_S(t)\leq 1$.}}}
\textcolor{black}{Note that the back-off process of each device is unchanged and its waiting rate is still $w$.}
When the device is in state (S), it will transit to state (I) after the device completes the transmission of one status update to the receiver.
For the scheme without preemption in service, it is obvious that the transition rate is $\mu$.
For the scheme with preemption in service, although a random number of status updates in service may be preempted, the service rate is $\mu$ for all status updates throughout the service period. Thus, the service period is memoryless in nature and is independent of the number of status updates that get preempted\cite{8469047}. Therefore, the corresponding transition rate is also $\mu$.
Fig.~\ref{fig:ctmc} illustrates the state transitions diagram of the CTMC $\{Q_n(t),t\geq 0\}$.
Let $\bs{\pi}_n\triangleq(\pi_{n,q})_{q\in\mathcal{Q}}$ be the stationary distribution of the CTMC of device $n$.

We adopt the AoI as our key performance metric to characterize the timeliness of the status information at the receiver for each device $n$, which is defined as the time elapsed since the most recently received status update was collected at device $n$\cite{6195689}.
In the following sections, we  first analyze  the average AoI and the average peak AoI of each device under the stationary distribution  $\bs{\pi}_n$ of the Markov process $\{Q_n(t),t\geq 0\}$ in Section~\ref{sec:aoi_analysis}.
The analytical results allow us to better understand the benefits of preemption in service for improving the AoI performance and the effects of the system parameters on the average AoI and average peak AoI.
Then, in Section~\ref{sec:mfg}, we formulate a noncooperative medium access game using which each device $n\in\mathcal{N}$ optimizes its waiting rate $w_n$ so as to minimize its average AoI or average peak AoI under an average energy cost constraint on channel sensing and packet transmitting.
By analyzing this game using a mean-field approximation, we can understand the strategic behavior of IoT devices in minimizing their AoI performance for the system with a large number of devices and obtain design insights for  practical ultra-dense IoT systems.


\begin{figure}[!t]
\begin{minipage}[h]{1\linewidth}
\centering
       \includegraphics[height=7cm]{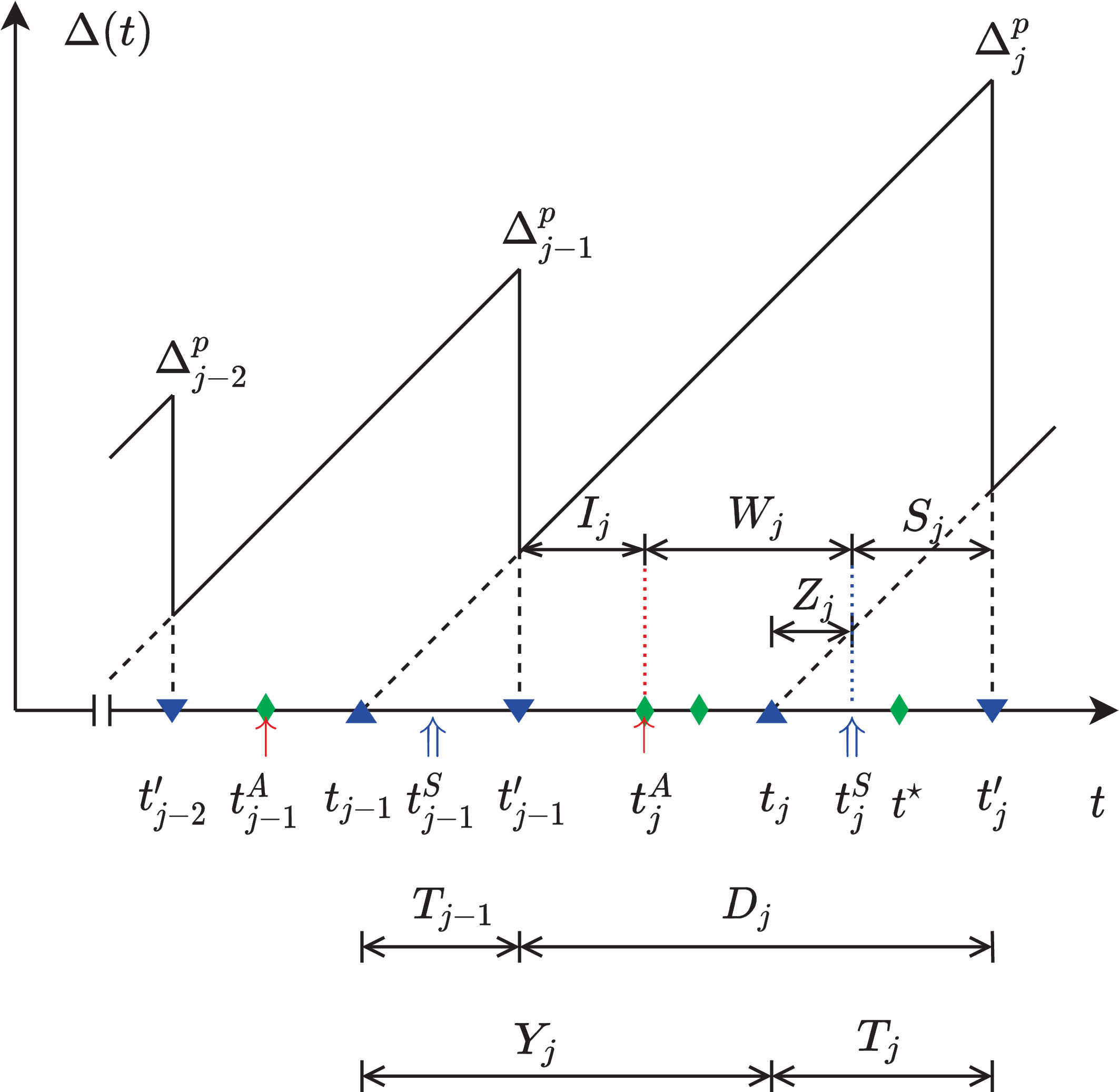}
\subcaption{}\label{fig:system_AoI_without_preemption}
\end{minipage}

\begin{minipage}[h]{1\linewidth}
\centering
        \includegraphics[height=7cm]{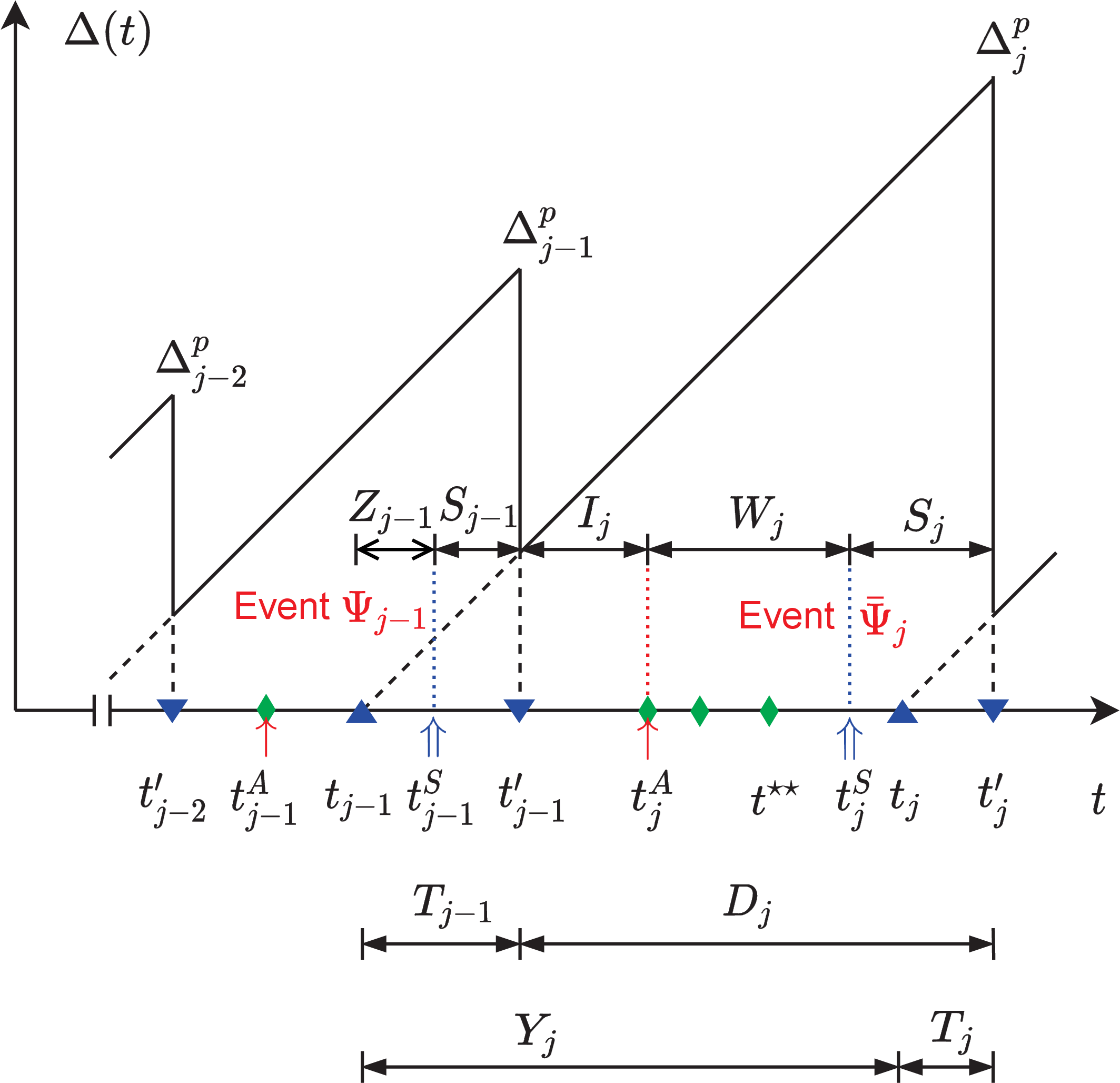}
\subcaption{}\label{fig:system_AoI_with_preemption}
\end{minipage}
 \caption{Illustration of the AoI evolution under two packet management schemes. On the time axis, the delivered status updates arrive at the device at times marked $\textcolor{blue}{\blacktriangle}$ and are delivered to its receiver at times marked $\textcolor{blue}{\blacktriangledown}$. The status updates that are discarded afterwards arrive at the device at  times marked $\textcolor{green}{\blacklozenge}$. \textcolor{black}{The device transits from state (I) to state (W) at times marked $\textcolor{red}{\uparrow}$ and transits from state (W) to state (S) at times marked $\textcolor{blue}{\Uparrow}$.} (a) The scheme without preemption in service. (b) The scheme with preemption in service.}
 \label{fig:system_AoI_evlution}

\end{figure}

 \section{Analysis of the Average AoI and the Average Peak AoI}\label{sec:aoi_analysis}
In this section, we derive the closed-form expressions of the average AoI and the average peak AoI under the stationary distribution $\bs{\pi}_n$ for the two packet management schemes, \textcolor{black}{based on a graphical approach\cite{6195689}. This is challenging because one must consider the effects of the existence of the random waiting period (due to the CSMA-type scheme) and packet preemption during the waiting period and/or the service period on the characterization of the complex temporal evolution of the AoI.
}

For the ease of presentation, we omit the subindex $n$ of the device hereinafter.
For each device, assuming that the freshest status update received at the receiver at time $t$ was generated at time $U(t)$, then the instantaneous AoI at $t$ for the device will be $\Delta(t)\triangleq t-U(t)$ \cite{6195689}.
We can see that the instantaneous AoI increases linearly with $t$ and is reset to a smaller value upon the delivery of a fresher status update, as illustrated in a sawtooth form in Fig.~\ref{fig:system_AoI_evlution}.

For each device, we define the time instants, $t_j$ and $t_j'$ as, respectively, the arrival time at the device and the delivery time at its receiver of the $j$-th delivered status update, for $j=1,2,\cdots$.
Thus, at time $t_j'$, the instantaneous AoI $\Delta(t_j')$ is reset to $T_j=t_j'-t_j$. We define $T_j$ as the service time of the $j$-th delivered status update.
\textcolor{black}{Note that the arrival at time $t_j$ is a fresh zero-age arrival.}
Let  $Y_j$ and $D_j$ be, respectively,  the inter-arrival time and the inter-departure time  between the $(j-1)$-th delivered update and the $j$-th delivered status updates, i.e., $Y_j = t_j - t_{j-1}$ and $D_j = t_j' - t_{j-1}'$.
For each  device, let $\Delta^p_j$ be the instantaneous peak AoI of the $j$-th delivered status update, which is defined as the instantaneous AoI immediately before the delivery of the $j$-th update\cite{7415972}, i.e.,
\begin{align}\label{eqn:defi_peak_aoi}
\Delta_j^p\triangleq\lim_{t\to [t_j']^{-}} \Delta(t).
\end{align}
From Fig.~\ref{fig:system_AoI_evlution}, we can see that the instantaneous peak AoI in \eqref{eqn:defi_peak_aoi} can also be expressed as:
\begin{align}\label{eqn:defi_peak_aoi_2}
\Delta_j^p=T_{j-1}+D_j= Y_j + T_{j}.
\end{align}

Now, we study the average AoI and the average peak AoI for each device under the stationary distribution $\bs{\pi}$ for the two considered packet management schemes. By \cite{8469047} and \cite{7415972}, the average AoI and the average peak AoI are respectively given by:
\begin{align}
&\bar{\Delta} \triangleq \lim_{\tau\to\infty}\frac{1}{\tau}\int_{0}^{\tau}\Delta(t) dt = \frac{\mathbb{E}[Y_jT_j]+\mathbb{E}[Y_j^2]/2}{\mathbb{E}[Y_j]},\label{eqn:defi_avg_aoi}\\
&\bar{\Delta}^p \triangleq \lim_{J\to\infty}\frac{1}{J}\sum_{j=1}^J\Delta_j^p = \mathbb{E}[\Delta_j^p] = \mathbb{E}[Y_{j}] + \mathbb{E}[T_j].\label{eqn:defi_avg_peak}
\end{align}

We next derive all the terms in \eqref{eqn:defi_avg_aoi} and \eqref{eqn:defi_avg_peak}.
We begin by introducing some additional notations and definitions.
Note that for each delivered status update, the device has gone through three time periods sequentially, corresponding to  states (I), (W), and (S) in Fig.~\ref{fig:ctmc}, respectively.
Let $t_j^A$ be the time instance at which the first status update arrives at the device after the delivery of the $(j-1)$-th status update.
Then, the device is in state (I) during the time period $(t_{j-1}',t_j^A]$ and moves to state (W) after time $t_j^A$.
Let $t_j^S$ be the time instance at which the device finds an idle channel after waiting for a backoff period in state (W).
Then, the device is in state (W) during the time period $(t_j^A,t_j^S]$ and in state (S) during the time period $(t_j^S,t_j']$.
Let $I_j$, $W_j$, and $S_j$ be, respectively, the idle duration, the waiting duration, and the service duration of the device for the $j$-th status update, i.e.,
$I_j=t_j^A-t_{j-1}'$, $W_j = t_j^S -t_j^A $, and $S_j = t_j' - t_j^S$.

Fig.~\ref{fig:system_AoI_without_preemption} and Fig.~\ref{fig:system_AoI_with_preemption} illustrate the evolution of the instantaneous AoI and the instantaneous peak AoI for the schemes without and with preemption, respectively.
Clearly, during the waiting period, any status update waiting for transmission will be substituted by a new arriving one and then be discarded.
During the service period, any arriving status update will be discarded, as seen from Fig.~\ref{fig:system_AoI_without_preemption}, and any arriving status update will preempt the one in service, as seen from Fig.~\ref{fig:system_AoI_with_preemption}.
For example, the status update arriving at time $t^{\star}$ during the $j$-th service period in Fig.~\ref{fig:system_AoI_without_preemption} is discarded, while this update (at time $t_j$ in Fig.~\ref{fig:system_AoI_with_preemption}) preempts the service of the status update that arrived at time $t^{\star\star}$ during the $j$-th waiting period in Fig.~\ref{fig:system_AoI_with_preemption}.

From Fig.~\ref{fig:system_AoI_evlution}, we can see that $D_j = I_j + W_j + S_j$ for both packet management schemes.
Recall that, under the stationary distribution $\bs{\pi}$, for each device, the idle duration $I_j$, the waiting duration $W_j$, and the service duration $S_j$ are mutually independent, exponentially distributed random variables with respective parameters $\lambda$, $w(1-\gamma \pi_S)$, and $\mu$.
For notational convenience, let $k\triangleq w(1-\gamma \pi_S)$.
 Then, we have
\begin{align}
&\mathbb{E}[D_j] = \frac{1}{\lambda} + \frac{1}{k} + \frac{1}{\mu},\label{eqn:E_D}\\
&\mathbb{E}[D_j^2]  = \frac{2}{\lambda^2} + \frac{2}{k^2} + \frac{2}{\mu^2} + \frac{2}{\lambda\mu} +\frac{2}{\lambda k} + \frac{2}{k\mu}.\label{eqn:E_D^2}
\end{align}
Since $T_{j-1}$ and $T_{j}$ are identically distributed, $T_{j-1}$ and $D_j$ are independent, and $T_{j}$ and $Y_j$ are independent, according to \eqref{eqn:defi_peak_aoi_2}, as done in \cite{8469047},  we can easily obtain that $\mathbb{E}[Y_j] = \mathbb{E}[D_j],$ $\mathbb{E}[Y_j^2] = \mathbb{E}[D_j^2]$, and $\mathbb{E}[Y_jT_j] =\mathbb{E}[Y_j]\mathbb{E}[T_j]$.
Now, in order to obtain $\bar{\Delta}$ and $\bar{\Delta}^p$ according to \eqref{eqn:defi_avg_aoi} and \eqref{eqn:defi_avg_peak}, we only need to compute the  remaining term $\mathbb{E}[T_j]$ and we will do so first for the scheme without preemption followed by the scheme with preemption.

\subsection{Scheme without Preemption in Service}
For the scheme without preemption in service, as seen from Fig.~\ref{fig:system_AoI_without_preemption}, $T_j$ is the time  that the $j$-th status update spends in waiting for the time interval $Z_j = t_j^S-t_j$ and in service throughout the service period $S_j$. This status update completes service and is not preempted during the interval $Z_j$.
Thus, $Z_j$ is the random time interval during which the device is in state (W) after the $j$-th status update arrives, and no new status updates arrive.
Note that $Z_j$ is independent of the fraction of the waiting period $W_j$ that had elapsed before $t_j$, due to  the memoryless property of the waiting period.
\textcolor{black}{Let $T_w$ be the random variable representing the time interval between the arrival time of a delivered status packet and the finish time of the corresponding waiting period, and let $T_{\lambda}$ be the random variable representing the time interval between the arrival time of a delivered status packet and the arrival time of next arriving status packet. For example, in Fig.~\ref{fig:system_AoI_without_preemption}, $T_w$ can represent the time interval between $t_j$ and $t_j^S$, and $T_{\lambda}$ can represent the time interval between $t_j$ and $t^*$.
Then, we can see that the distribution of $Z_j$ is that of the time interval $T_w$, conditioned on  $T_w < T_{\lambda}$.}
The probability density function of $Z_j$ can be derived as:
\begin{align}\label{eqn:pdf_z_wop}
&f_{Z}(t)= \lim_{dt\to 0} \frac{\Pr[t< T_w\leq t+dt\mid T_w<T_{\lambda}]}{dt}\nonumber\\
&= \lim_{dt\to 0} \frac{\Pr[t< T_w\leq t+dt]\Pr[T_w<T_{\lambda}\mid t< T_w\leq t+dt]}{dt \Pr[T_w<T_{\lambda}]}\nonumber\\
& = \frac{f_{T_w}(t)\Pr[T_{\lambda}>t]}{\Pr[T_{\lambda}>T_w]}.
\end{align}
Note that $T_w$ and $T_{\lambda}$ are exponential distributed with parameters $k$ and  $\lambda$, respectively, due to the memoryless property of the waiting times and the status update arrival times.
Thus, we have $f_{T_w}(t) = k e^{-k t}$, where $t>0$, $\Pr[T_{\lambda}>t] = e^{-\lambda t}$, and $\Pr[T_{\lambda}>T_w] =\int_{0}^{\infty}f_{T_w}(t)\Pr[T_{\lambda}>t]dt = \frac{k}{k+\lambda}$. By \eqref{eqn:pdf_z_wop}, we obtain $f_{Z}(t) = (k+\lambda)e^{-(k+\lambda)t}$ and $\mathbb{E}[Z_j] = \frac{1}{k+\lambda}$.
Therefore, we have
\begin{align}
\mathbb{E}[T_j] = \mathbb{E}[Z_j] + \mathbb{E}[S_j] = \frac{1}{k+\lambda} + \frac{1}{\mu}. \label{eqn:E_Tj_wop}
\end{align}
\subsection{Scheme with Preemption in Service}
For the scheme with preemption in service, the service time $T_j$ depends on whether there is any status update arrival during the service period $S_j$, as can be seen from the cases for the $(j-1)$-th and the $j$-th delivered status updates in Fig.~\ref{fig:system_AoI_with_preemption}.
Let $\Psi_j$ be the event that no status update arrivals occur during the service period $S_j$ and $\bar{\Psi}_j$ be the complement event of $\Psi_j$.
Due to the memoryless property of the status update arrival times and the service times, the event $\Psi_j$ is equivalent to  the event  that the time interval from $t_j^S$ to the first occurrence of a status update arrival, $T_\lambda$, is greater than the service period $S$. Then, we have
\begin{align}
&\Pr[\Psi_j]= \Pr[T_{\lambda}>S] =  \int_{0}^{\infty} \Pr[T_{\lambda}>t]f_{S}(t) dt\nonumber\\
&\hspace{31mm}= \int_{0}^{\infty} e^{-\lambda t} \mu e^{-\mu t} dt = \frac{\mu}{\lambda + \mu},\label{eqn:prob_psi}\\
&\Pr[\bar{\Psi}_j] = 1 - \Pr[\Psi_j] = \frac{\lambda}{\lambda + \mu}.\label{eqn:prob_bar_psi}
\end{align}

Now, we derive the conditional distributions $f_{T}(t\mid\Psi_j)$ and $f_{T}(t\mid\bar{\Psi}_j)$ of the service time $T_j$ given the events $\Psi_j$ and $\bar{\Psi}_j$, respectively.
Conditioned on the event $\Psi_j$,  $T_j$ is the time  that the $j$-th status update spends in waiting for the time interval $Z_j = t_j^S-t_j$ and in service throughout the service period $S_j$. This is the same to  that for the scheme without preemption in service.
Note that $Z_j$ and $S_j$ are independent, exponential random variables with parameters $k$ and $\mu$. Then, we can obtain the probability distribution of \textcolor{black}{$T_j = Z_j + S_j$}, given by:
\begin{align}
f_T(t) = \frac{k\mu}{\mu-k} \left(e^{-kt}-e^{-\mu t}\right).
\end{align}
The $j$-th status update completes its service and thus, no new status update arrivals will occur during $T_j$.
Therefore, under the event $\Psi_j$, the conditional distribution of $T_j$ is that of the service time $T_j$, conditioned on $T_j$ being smaller than the time interval from $t_j$ to the next status update arrival, $T_{\lambda}$.
Then, the conditional distribution of the service time $T_j$ is given by:
\begin{align}\label{eqn:pdf_T_wp_case1}
&f_{T}(t\mid\Psi_j)= \lim_{dt\to 0} \frac{\Pr[t< T\leq t+dt\mid T<T_{\lambda}]}{dt}\nonumber\\
&= \lim_{dt\to 0} \frac{\Pr[t< T\leq t+dt]\Pr[T<T_{\lambda}\mid t< T\leq t+dt]}{dt \Pr[T<T_{\lambda}]}\nonumber\\
& = \frac{f_{T}(t)\Pr[T_{\lambda}>t]}{\int_{0}^{\infty}f_{T}(t)\Pr[T_{\lambda}>t]dt}\nonumber\\
&= \frac{(\lambda+k)(\lambda + \mu)}{\mu-k} \left(e^{-(\lambda + k)t}-e^{-(\lambda + \mu)t}\right),
\end{align}
with the conditional expectation:
\begin{align}
\mathbb{E}[T_j\mid\Psi_j] =  \frac{1}{\lambda + k} + \frac{1}{\lambda + \mu}.\label{eqn:e_t_psi}
\end{align}
Conditioned on the complement event $\bar{\Psi}_j$, $T_j$ is equivalent to the service duration $S_j$.
\textcolor{black}{Note that the delivered $j$-th packet did not spend time for waiting as this packet arrived during the service period and preempted an undergoing service. Thus, different from the case for the event $\Psi_j$,  there does not exist the interval $Z_j$ conditioned on the event $\bar{\Psi}_j$.}
Thus, the conditional distribution of $T_j$ given the event $\bar{\Psi}_j$ is that of the service duration $S_j$, conditioned on $S_j$ being smaller than $T_{\lambda}$. Similar to \eqref{eqn:pdf_T_wp_case1}, the conditional distribution $f_{T}(t\mid\bar{\Psi}_j)$ can be derived as:
\begin{align}\label{eqn:pdf_T_wp_case2}
f_{T}(t\mid\bar{\Psi}_j) = \frac{f_{S}(t)\Pr[T_{\lambda}>t]}{\int_{0}^{\infty}f_{S}(t)\Pr[T_{\lambda}>t]dt}
= \frac{1}{\lambda + \mu} e^{-(\lambda + \mu)t},
\end{align}
with the conditional expectation:
\begin{align}
\mathbb{E}[T_j\mid\bar{\Psi}_j] =  \frac{1}{\lambda + \mu}.\label{eqn:e_t_bar_psi}
\end{align}
Based on the probabilities for the events $\Psi_j$ and $\bar{\Psi}_j$ calculated in \eqref{eqn:prob_psi} and \eqref{eqn:prob_bar_psi}, as well as the conditional expectations calculated in \eqref{eqn:e_t_psi} and \eqref{eqn:e_t_bar_psi}, we now derive the expected value of $T_j$:
\begin{align}
\mathbb{E}[T_j] &= \Pr[\Psi_j]\mathbb{E}[T_j\mid\Psi_j] + \Pr[\bar{\Psi}_j]\mathbb{E}[T_j\mid\bar{\Psi}_j]\nonumber\\
& =\frac{\mu}{\lambda + \mu}\left(\frac{1}{\lambda + k} + \frac{1}{\lambda + \mu}\right) + \frac{\lambda}{\lambda + \mu}\frac{1}{\lambda + \mu}\nonumber\\
& = \frac{1}{\lambda + \mu} \left(1 + \frac{\mu}{\lambda + k}\right).\label{eqn:E_Tj_wp}
\end{align}

Finally, by applying \eqref{eqn:E_D}, \eqref{eqn:E_D^2}, \eqref{eqn:E_Tj_wop}, and \eqref{eqn:E_Tj_wp} into \eqref{eqn:defi_avg_aoi} and \eqref{eqn:defi_avg_peak}, we can derive the average AoI  and the average peak AoI for the two packet management schemes with and without preemption in service, as summarized in the following theorem, \emph{whose proof is based on the previous analysis}.
\begin{theorem} \label{theorem:aoi}
Under the stationary distribution $\bs{\pi}$, for each device, the average AoI and the average peak AoI are given as follows.
For the packet management scheme without preemption in service, we have
\begin{align}
&\bar{\Delta}_{WOP} =  \frac{1}{\lambda} + \frac{1}{k} + \frac{2}{\mu} + \frac{1}{\lambda + k } - \frac{\lambda + k + \mu}{\lambda k + k\mu + \lambda\mu},\label{eqn:avg_aoi_wop}\\
&\bar{\Delta}^p_{WOP} = \frac{1}{\lambda} + \frac{1}{k} + \frac{2}{\mu} + \frac{1}{\lambda + k },\label{eqn:avg_peak_aoi_wop}
\end{align}
and for the packet management scheme with preemption in service, we have
\begin{align}
&\bar{\Delta}_{WP} =  \frac{1}{\lambda} + \frac{1}{k} + \frac{1}{\mu} + \frac{1}{\lambda + \mu} \left(1 + \frac{\mu}{\lambda + k}\right) \nonumber\\
&\hspace{12mm}- \frac{\lambda + k + \mu}{\lambda k + k\mu + \lambda\mu},\label{eqn:avg_aoi_wp}\\
&\bar{\Delta}^p_{WP} = \frac{1}{\lambda} + \frac{1}{k} + \frac{1}{\mu} +\frac{1}{\lambda + \mu} \left(1 + \frac{\mu}{\lambda + k}\right),\label{eqn:avg_peak_aoi_wp}
\end{align}
where $k= w (1-\gamma \pi_S)$.
\end{theorem}

\textcolor{black}{Note that the expressions of the average AoI and the average peak AoI are derived by assuming that the system is in the steady state $\bs{\pi}$ with the waiting rate $k=w(1-\gamma\pi_S)$.}
The average AoI and the average peak AoI functions derived in Theorem~\ref{theorem:aoi} depend on the stationary distribution $\bs{\pi}$ only through the fraction of devices in state (S) $\pi_S$.
From Theorem~\ref{theorem:aoi}, we can verify that $\bar{\Delta}_{WP}< \bar{\Delta}_{WOP}$ and $\bar{\Delta}_{WP}^p< \bar{\Delta}_{WOP}^p$, i.e.,
the packet scheme with preemption in service always achieves a smaller average AoI and a smaller average peak AoI, compared to the scheme without preemption in service.
Moreover, we can see that in the limiting case with the rate $k$ going to infinity (i.e., the device transmits immediately upon a new status update arrives and there always exist available channels for transmission), the average AoI under the two packet management schemes will be given by
\begin{align}
&\lim_{k\to\infty} \bar{\Delta}_{WOP} =  \frac{1}{\lambda} + \frac{2}{\mu} - \frac{1}{\lambda + \mu},\label{eqn:avg_aoi_wop_limit} \\
&\lim_{k\to\infty} \bar{\Delta}_{WP} =  \frac{1}{\lambda} + \frac{1}{\mu}.\label{eqn:avg_aoi_wp_limit}
\end{align}
Note that \eqref{eqn:avg_aoi_wop_limit} is the same to the average AoI in an FCFS M/M/1/1 queue \cite{7415972} and  \eqref{eqn:avg_aoi_wp_limit} is the same to the average AoI in an LCFS M/M/1/1 queue \cite{8469047}.
Theorem~\ref{theorem:aoi} provides a rigorous analytical characterization of the average AoI and the average peak AoI for IoTs with random status packets arrivals, under a CSMA-type scheme.
Given the derived closed-form expressions in Theorem~\ref{theorem:aoi}, we next investigate how to optimize the channel access strategy (i.e., the waiting rate) so as to minimize the average AoI and the average peak AoI of each device \textcolor{black}{under an average energy cost constraint for each device}.

\section{A Mean-Field Game Framework for AoI Minimization}\label{sec:mfg}
In this section, we aim at designing the optimal waiting rate of each device so that  its average AoI or its average peak AoI is minimized under the associated stringent energy constraint.

\textcolor{black}{In our model, for each device, there will incur some energy cost for channel sensing during state (W) and for status update transmission during state (S). Recall that  $C_s$ and $C_t$ are the channel sensing cost per unit time and the transmission cost per unit time for each device, respectively.}
Then, under the stationary distribution $\bs{\pi}_n$, based on Fig.~\ref{fig:system_AoI_evlution} and by using the renewal reward theorem\cite{stewart2009probability},
we can derive the average energy cost per unit time of  device $n$
for channel sensing and packet transmitting as:
\begin{align}
\bar{C}(w_n,\bs{\pi}_n) = \dfrac{\mathbb{E}[W_j]C_s+\mathbb{E}[S_j]C_t}{\mathbb{E}[D_j]}
=  \frac{\frac{C_s}{w_n(1-\gamma \pi_{n,S})}+\frac{C_t}{\mu}}{\frac{1}{\lambda}+\frac{1}{w_n(1-\gamma \pi_{n,S})}+\frac{1}{\mu}}.
\end{align}
Note that $\bs{\pi}_n$ depends on the waiting rates of all devices $\bs{w}\triangleq (w_m)_{m\in\mathcal{N}}$.
Given the waiting rates chosen by all other devices $\bs{w}_{-n}\triangleq \bs{w}\setminus w_n$, each device $n\in\mathcal{N}$ seeks to find its optimal waiting rate $w_n$ by solving the following optimization problem:
\begin{subequations}\label{eqn:problem_original}
\begin{align}
&\min_{w_n} g(w_n,\bs{\pi}_n(w_n,\bs{w}_{-n})),\label{eqn:obj_ori}\\
&~\text{s.t.}~\bar{C}(w_n,\bs{\pi}_n(w_n,\bs{w}_{-n}))\leq \hat{C},\label{eqn:cons_ori}
\end{align}
\end{subequations}
where $g(\cdot)\in\{\bar{\Delta}_{WOP},\bar{\Delta}^p_{WOP},\bar{\Delta}_{WP},\bar{\Delta}^p_{WP}\}$ is a generic AoI function corresponding to the average AoI and the average peak AoI under the two packet management schemes  derived in Theorem~\ref{theorem:aoi}.
The problem in \eqref{eqn:problem_original} is a noncooperative game\cite{han2019game}, as the objective function in \eqref{eqn:obj_ori} and the constraint in \eqref{eqn:cons_ori} are coupled by the actions (i.e., the waiting rates) of all devices.
In general, finding the equilibrium of the game in \eqref{eqn:problem_original} is computationally prohibitive for an ultra-dense IoT system with a large $N$\cite{nisan_roughgarden_tardos_vazirani_2007}.
Moreover, by the transition diagram of $\{Q_n(t),t\geq 0\}$ in Fig.~\ref{fig:ctmc}, we can see that it is a non-homogeneous CTMC with time-varying transition rates, and thus, it is generally impossible to derive its stationary distribution \cite{6239591}.

To tackle the aforementioned challenges,  we  investigate the considered IoT monitoring system in a \emph{mean-field} regime when the number of devices grows large, by using a mean field approximation \cite{10.1145/2825236.2825241,10.1145/3084454,10.1145/2964791.2901463,10.1145/3323679.3326498}.
In the mean-field regime, each individual device would have a negligible impact on the stochastic system $\bs{X}$.
This enables us to characterize the game equilibrium of \eqref{eqn:problem_original} via the interaction between a typical device and the ``average'' of all other devices (i.e., the mean-field), instead of focusing  on the more complex interactions among all the devices.
Specifically, in the mean-field regime, we can replace the complex stochastic system $\bs{X}$ by a much simpler deterministic dynamic system, and then  obtain the explicit approximate expression of the stationary distribution $\bs{\pi}$.
Such a mean-field regime not only provides analytical tractability, but is also practical for emerging ultra-dense IoT systems.

 \subsection{A Mean Field Game Formulation}
 In our MFG formulation, we focus on the ultra-dense in which the number of devices $N$ goes to infinity  and $\gamma$ is a constant.

First, for a finite $N$, assuming that all devices use the same waiting rate $w$, based on Fig.~\ref{fig:ctmc}, we can derive the transitions of the process $\{\bs{X}(t),t\geq 0\}$:
\begin{equation}\label{eqn:transition_X}
 \left\{\left.\begin{aligned}
        &\bs{X} \mapsto \bs{X} + \frac{1}{N}(-1,1,0)\quad\text{at~rate}~N\lambda X_I,\\
        &\bs{X} \mapsto \bs{X} + \frac{1}{N}(0,-1,1)\quad\text{at~rate}~Nw(1-\gamma X_S) X_W,\\
        &\bs{X} \mapsto \bs{X} + \frac{1}{N}(1,0,-1)\quad\text{at~rate}~N\mu X_S.
       \end{aligned}
 \right.\right.
\end{equation}
Note that, as all devices are exchangeable, we have $\mathbb{E}[X_q(t)] = \frac{1}{N}\sum_{n=1}^N \Pr[Q_n(t)=q] = \Pr[Q_n(t)=q]$, for each $q\in\mathcal{Q}$ \cite{10.1145/3084454}.
Let $\bs{\pi}$ be the stationary distribution of  $\{\bs{X}(t)\}$, which is also the same to $\bs{\pi}_n$ for each device $n$.
From \eqref{eqn:transition_X}, we observe that it is still infeasible to compute the stationary distribution $\bs{\pi}$ of the CTMC $\bs{X}$, given that the state space $\{0,1/N,2/N,\cdots,1\}^3$  can be very large and the time-varying transition rates are nonlinear functions of the states.

The form of the transitions in \eqref{eqn:transition_X} indicates that $\bs{X}$ belongs to the class of density-dependent population processes \cite{10.1145/2964791.2901463,10.1145/3084454,doi:10.1137/1.9781611970333} and  the corresponding mean-field model of this CTMC can be characterized by the following ordinary differential equation (ODE):
\begin{align}
\dot{\bs{x}} = f(\bs{x}),\label{eqn:ode}
\end{align}
where $\bs{x} \triangleq (x_{I},x_{W},x_{S})$ and  $f(\bs{x}) \triangleq \lim\limits_{dt\to 0} \left(\mathbb{E}[\bs{X}(t+dt)-\bs{X}(t)\mid \bs{X}(t)=\bs{x}]\right)/dt$ is the drift.
By \eqref{eqn:transition_X}, we further have
\begin{equation}\label{eqn:mean-field}
 \left\{
 \left.\begin{aligned}
        &\dot{x}_I = -\lambda x_I + \mu x_S,\\
        &\dot{x}_W =  \lambda x_I -  w(1-\gamma x_S) x_W,\\
        &\dot{x}_S = w(1-\gamma x_S) x_W - \mu x_S.
       \end{aligned}
 \right.\right.
\end{equation}
We can see that $f(\bs{x})dt$ is the expected variation of the CTMC $\bs{X}$ that would start from the state $\bs{X}$ in a small time interval $dt$.
Let $\bs{x}^*$ be an equilibrium point of the ODE in \eqref{eqn:mean-field}.
\textcolor{black}{Here, $\bs{x}^*$ represents the steady state of the mean-field system. In other words, if the dynamic system starts with the equilibrium point $\bs{x}(0) =\bs{x}^*$, then the system state $\bs{x}(t)$ always remains at $\bs{x}^*$.}
Next, we show that the above mean-field approximation is accurate for the considered system.

\begin{theorem}\label{theorem:mean-field}
Assuming that all devices use the same waiting rate $w$, then the equilibrium point $\bs{x}^*=(x_I^*,x_W^*,x_S^*)$ of the mean-field model in \eqref{eqn:mean-field} is unique, and satisfies
\begin{subequations}\label{eqn:mean-field-solution}
\begin{align}
x_I^* =& \frac{\mu}{\lambda}x_s^*,\\
x_W^* =& \frac{\mu x_S^*}{w(1-\gamma x_S^*)},\\
x_S^* =& \frac{1}{2w\gamma(\lambda+\mu)}\Big(w(\lambda + \mu + \lambda \gamma) + \lambda\mu \nonumber\\
&- \sqrt{(w(\lambda + \mu + \lambda \gamma) + \lambda\mu)^2 -4\lambda (\lambda + \mu)\gamma w^2}\Big).
       \end{align}
\end{subequations}
Moreover, as $N$ goes to infinity, the stationary distribution $\bs{\pi}$, the AoI cost $g(w,\bs{\pi})$, and the energy cost $\bar{C}(w,\bs{\pi})$ converge, respectively, to $\bs{x}^*$,  $g(w,\bs{x}^*)$, and $\bar{C}(w,\bs{x}^*)$, with the following rates of convergence:
\begin{subequations}\label{eqn:rates_of_convergence}
\begin{align}
&\left|\mathbb{E}[\bs{\pi}] - \bs{x}^*\right| = O(\frac{1}{N}),\\
&\left|\mathbb{E}[g(w,\bs{\pi})] - g(w,\bs{x}^*)\right| = O(\frac{1}{N}),\\
&\left|\mathbb{E}[\bar{C}(w,\bs{\pi})] - \bar{C}(w,\bs{x}^*)\right|= O(\frac{1}{N}),
\end{align}
\end{subequations}
where the expectation is taken over the stationary distribution $\bs{\pi}$.
\end{theorem}
\begin{IEEEproof}
  See Appendix~\ref{app:mean-field}.
\end{IEEEproof}

\textcolor{black}{From Theorem~\ref{theorem:mean-field}, by the L'H\^{o}pital's rule and after some calculations, we can see that if $w\to 0$, then $x_I^*\to 0$, $x_W^*\to 0$, and $x_S^*\to 0$; and if $w\to \infty$, then $x_I^*\to \frac{\mu}{\lambda+\mu}$, $x_W^*\to 0$, and $x_S^*\to \frac{\lambda}{\lambda+\mu}$.}
According to Theorem~\ref{theorem:mean-field}, we can now approximate $\bs{\pi}$ with the equilibrium point $\bs{x}^*$ at the mean-field limit to solve the problem in \eqref{eqn:problem_original}.
\textcolor{black}{Later, in the simulations, we shall show that the approximation is very accurate, even for a small value of $N$.}
For a given $\bs{x}^*$, each device seeks to find the optimal waiting rate $w$ by solving the problem in the mean-field limit:
\begin{subequations}\label{eqn:problem_mean_field}
\begin{align}
&\min_{w} g(w,\bs{x}^*),\\
&~\text{s.t.}~\bar{C}(w,\bs{x}^*)\leq \hat{C}.\label{eqn:problem_mean_field-constraint}
\end{align}
\end{subequations}

Through the MFG in \eqref{eqn:problem_mean_field}, each device is viewed to interact with the mean-field of the system, instead of the explicit actions of all other devices.
\textcolor{black}{Here, we adopt a time-scale separation such as the devices adjust their waiting strategies in a slower time scale than the convergence of the mean-field model, as in \cite{10.1145/2825236.2825241} and \cite{10.1145/3323679.3326498}. Under this assumption, when it is the time for the devices to adjust their waiting rates, all devices can measure the fraction of busy channels accurately. Then, each device can compute its AoI function and its average energy cost function in \eqref{eqn:problem_mean_field-constraint}.}
Note that, for a given  equilibrium point $\bs{x}$, all devices will choose the same waiting rate $w$, as the AoI function $g(\cdot)$ and the energy cost function $\bar{C}(\cdot)$ are the same for them.
Let  $\textcolor{black}{\mathcal{L}_w}: \bs{x} \rightarrow w$ be the mapping from $\bs{x}$ to the optimal solution $w$ of the problem in \eqref{eqn:problem_mean_field} and let $\textcolor{black}{\mathcal{L}_{MF}}: w \rightarrow \bs{x}$ be the mapping from the waiting rate $w$ to the equilibrium point $\bs{x}$ of the mean field model in  \eqref{eqn:mean-field}, as given in  \eqref{eqn:mean-field-solution}.
Then, following\cite{10.1145/2825236.2825241} and\cite{doncel_gast_gaujal_2020}, we define  a strategy $w^*$ as a \emph{symmetric mean field equilibrium (MFE)}, if
\begin{align}\label{eqn:MFE}
w^* = \textcolor{black}{\mathcal{L}_w(\mathcal{L}_{MF}(w^*))}.
\end{align}
Under the MFE, all devices will use the waiting rate $w^*$ and no device can achieve better AoI performance by unilaterally deviating from its strategy in the mean-field limit.
Next, we analyze the existence and uniqueness of the MFE in \eqref{eqn:MFE}.

\subsection{Mean Field Equilibrium Analysis for the MFG}
First, we denote $\theta\triangleq \gamma x_S$. Note that $\theta$ indicates the probability that any channel is sensed busy in the mean-field limit, 
i.e., the fraction of busy channels.
Then, we characterize the closed-form solution for problem  \eqref{eqn:problem_mean_field} under a given $\bs{x}$, in the following lemma.
\begin{lemma}\label{lemma:optimal_w_for_given_theta}
Under a given $\bs{x}$ such that $0<\theta<1$, for any $g(\cdot)\in\{\bar{\Delta}_{WOP},\bar{\Delta}^p_{WOP},\bar{\Delta}_{WP},\bar{\Delta}^p_{WP}\}$, the optimal waiting rate obtained by solving \eqref{eqn:problem_mean_field} is the same and is given by:
\begin{align}\label{eqn:opt_w}
w^* = \begin{cases}&\dfrac{\hat{C}/(1-\theta)}{C_s/(1-\theta) + C_t/\mu  - \left(1/\lambda + 1/\mu\right)\hat{C}},  \\
&\hspace{18mm} \text{if}~\frac{C_s}{1-\theta} + \frac{C_t}{\mu}  > \left(\frac{1}{\lambda} + \frac{1}{\mu}\right)\hat{C}, \\
            &+\infty,\hspace{10mm}\text{otherwise}.
  \end{cases}.
\end{align}
\end{lemma}
\begin{IEEEproof}
First, we show that under a given $\bs{x}$, any $g(\cdot)\in\{\bar{\Delta}_{WOP},\bar{\Delta}^p_{WOP},\bar{\Delta}_{WP},\bar{\Delta}^p_{WP}\}$ is decreasing with $w$.
For the average peak AoI $\bar{\Delta}_{WOP}^p$ and $\bar{\Delta}_{WP}^p$, the monotonicity property can be easily seen from \eqref{eqn:avg_peak_aoi_wop} and \eqref{eqn:avg_peak_aoi_wp}.
The monotonicity property of the average AoI $\bar{\Delta}_{WOP}$ in \eqref{eqn:avg_aoi_wop} and $\bar{\Delta}_{WP}$ in \eqref{eqn:avg_aoi_wp} can be obtained by taking the corresponding derivatives with respect to $k=w(1-\theta)$:
\begin{align}
&\frac{\partial \bar{\Delta}_{WOP}}{\partial k} = -\frac{1}{(\lambda+k)^2} -\frac{\lambda\mu(k^2+2k\mu+2\lambda k +\lambda\mu)}{k^2(\lambda k + k\mu + \lambda\mu)^2}<0,\\
&\frac{\partial \bar{\Delta}_{WP}}{\partial k} = -\frac{\mu}{(\lambda+\mu)(\lambda+k)^2} -\frac{\lambda\mu(k^2+2k\mu+2\lambda k +\lambda\mu)}{k^2(\lambda k + k\mu + \lambda\mu)^2}\nonumber\\
&\hspace{12mm}<0.
\end{align}
Then, for the constraint in \eqref{eqn:problem_mean_field-constraint}, after some calculations, we can rewrite it as:
\begin{align}
\left( \frac{C_s}{1-\theta} + \frac{C_t}{\mu}  - \left(\frac{1}{\lambda} + \frac{1}{\mu}\right)\hat{C} \right) w \leq \frac{\hat{C}}{1-\theta}
\end{align}
Therefore, we can obtain the optimal $w$ in \eqref{eqn:opt_w}, which minimizes any $g(\cdot)\in\{\bar{\Delta}_{WOP},\bar{\Delta}^p_{WOP},\allowbreak\bar{\Delta}_{WP},\allowbreak\bar{\Delta}^p_{WP}\}$. We complete the proof.
\end{IEEEproof}

Lemma~\ref{lemma:optimal_w_for_given_theta} indicates that, under a given $\theta$, the optimal waiting rate $w$ depends only on the energy constraint \eqref{eqn:problem_mean_field-constraint} and each device will fully utilize its available energy. The reason is that the average AoI and the average peak AoI functions derived in Theorem~\ref{theorem:aoi} are all decreasing with $w$.
Moreover, when $\frac{C_s}{1-\theta} + \frac{C_t}{\mu}  \leq \left(\frac{1}{\lambda} + \frac{1}{\mu}\right)\hat{C}$, we can see that the optimal waiting rate of each device is $w=\infty$, i.e., each device does not need to wait for a certain waiting period and should start its transmission whenever there is a new status update arrival and an idle channel is sensed. This indicates that, each device has sufficient energy for channel sensing and status update transmission in this case.

Based on the equilibrium point $\bs{x}^*$ for a given waiting rate $w$ characterized in Theorem~\ref{theorem:mean-field} and  the optimal waiting rate $w$ for a given equilibrium point $\bs{x}^*$ characterized in Lemma~\ref{lemma:optimal_w_for_given_theta}, we now characterize the existence and uniqueness of the \textcolor{black}{MFE} in \eqref{eqn:MFE} in the following theorem.

\begin{theorem}\label{theorem:MFE}
The existence and uniqueness of the \textcolor{black}{MFE} depends on the following three cases.
\begin{enumerate}
\item If $\frac{C_s}{\max\{0,1-\frac{\gamma\lambda}{\lambda+\mu}\}} + \frac{C_t}{\mu}  \leq (\frac{1}{\lambda} + \frac{1}{\mu})\hat{C}$,
then,  $w^*=\infty$ is the unique \textcolor{black}{MFE}, i.e., each device should transmit its status update immediately without waiting, whenever there is a new status update arrival and any idle channel is sensed.
\item If $\frac{C_s}{1-\theta^*} + \frac{C_t}{\mu}  > (\frac{1}{\lambda} + \frac{1}{\mu})\hat{C}$,
where $\theta^* =\frac{1}{2C_t}(\gamma\hat{C} + \mu C_s + C_t - \sqrt{(\gamma\hat{C} + \mu C_s + C_t)^2-4\gamma C_t\hat{C}} )$,
then the unique \textcolor{black}{MFE} is given by:
\begin{align}\label{eqn:case2-MNE}
w^* = \frac{ \hat{C}/(1-\theta^*)} {C_s/(1-\theta^*) + C_t/\mu  - \left(1/\lambda + 1/\mu\right)\hat{C}}.
\end{align}
\item
For the remaining case,
i.e., if $\frac{C_s}{1-\theta^*} + \frac{C_t}{\mu}  \leq (\frac{1}{\lambda} + \frac{1}{\mu})\hat{C} <\frac{C_s}{\max\{0,1-\frac{\gamma\lambda}{\lambda+\mu}\}} + \frac{C_t}{\mu}$,
the MFE does not exist and each device switches its waiting rate between $w=\infty$ and
\textcolor{black}
{\begin{align}\label{eqn:case3_w}
w = \frac{\hat{C}}{C_s + (C_t/\mu  - \left(1/\lambda + 1/\mu\right)\hat{C})\max\{0,1-\frac{\gamma\lambda}{\lambda+\mu}\}}.
\end{align}
}
\end{enumerate}
\end{theorem}
\begin{IEEEproof}
  See Appendix~\ref{app:MFE}.
\end{IEEEproof}

\textcolor{black}{From Theorem~\ref{theorem:MFE}, under Case 1) and Case 2), the mean-field solution is a policy with a fixed waiting rate.}
For Case 1), it can be seen that, in order to satisfy its condition, $\frac{\gamma\lambda}{\lambda+\mu}<1$ must hold. This indicates that, each device can transmit without waiting only if there are sufficient communication resources (i.e., $\gamma = N/M$ is small) and the channel utilization load $\lambda/\mu$ is low.
Moreover, by Theorem~\ref{theorem:aoi}, we can easily see that the average AoI and the average peak AoI decrease, as the status update arrival rate $\lambda$ increases, under the MFE in Case 1).
For Case 2), it can be seen that, if the stated condition is satisfied, then the fraction of busy channels $\theta^*$  in the mean-field limit is independent of the arrival rate $\lambda$, and thus, the waiting rate $w^*$ in \eqref{eqn:case2-MNE} decreases with $\lambda$.
In other words, when the status update arrival rate increases, each device must wait for a longer period before transmitting its status update without violating the energy constraint.
Then,  based on the closed-form expressions derived in Theorem~\ref{theorem:aoi}, we see that,  for both packet management schemes, the average AoI and the average peak AoI do not necessarily decrease with the arrival rate $\lambda$, under the MFE in Case 2). This contradicts our intuition that the average AoI and the average peak should always decrease with the arrival rate, for the system in which preemption is always allowed during waiting  and can be allowed in transmission, as seen from \cite{8469047} and \cite{7415972}.
\textcolor{black}{For Case 3), if the devices choose $\tilde{w}=\infty$, the fraction of devices in state (S) will converge to \textcolor{black}{$\tilde{x}_S =\frac{\lambda}{\lambda + \mu}$} in the mean-field limit.
After measuring the fraction of busy channels is $\tilde{\theta} = \gamma\tilde{x}_S$, all devices will change their waiting rates to $w$ in \eqref{eqn:case3_w}.
Then, after measuring the fraction of busy channels under $w$, all devices will switch their waiting rates back to $w=\infty$. Thus, there does not exist an equilibrium solution of the MFG under Case 3).}

Next, we study the convergence of the MFE for Case 1) and Case 2) in Theorem~\ref{theorem:MFE}.
\begin{proposition}\label{prop:convergence}
For Case 1) in Theorem~\ref{theorem:MFE}, if its condition is satisfied, then the MFG converges to the MFE $w^*=\infty$ starting from any initial condition.
For Case 2) in Theorem~\ref{theorem:MFE}, the MFG converges to the MFE in \eqref{eqn:case2-MNE} starting from any initial condition, if both the condition for Case 2) and the following condition are satisfied:
\begin{align}\label{eqn:convergence-case2}
\frac{\gamma \hat{C}}{\mu B^2} \left|\frac{C_t}{\mu} - \left(\frac{1}{\lambda} + \frac{1}{\mu}\right)\hat{C}\right| < 1,
\end{align}
where $B=\min\{C_s,C_s + \frac{C_t}{\mu}  - (\frac{1}{\lambda} + \frac{1}{\mu})\hat{C}\}$.
\end{proposition}
\begin{IEEEproof}
See Appendix~\ref{app:convergence}.
\end{IEEEproof}

\begin{figure}[!ht]
\begin{centering}
\includegraphics[scale=.45]{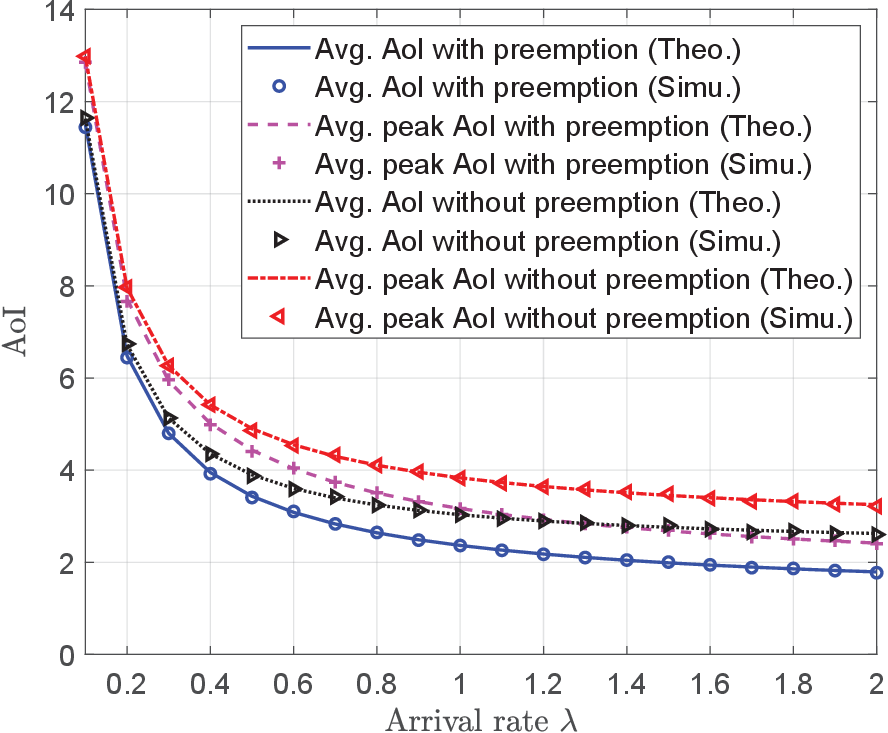}
 \caption{Average AoI and average peak AoI versus the arrival rate $\lambda$ under the packet management schemes with and without preemption in service. $\mu=1$ and $k=2$.
} \label{fig:AoI_expression_verify}
\end{centering}
\end{figure}

Proposition~\ref{prop:convergence} provides conditions under which the MFG will converge to an MFE starting from any initial system states.
Note that the condition in \eqref{eqn:convergence-case2} for Case 2) is a sufficient condition for the convergence to the MFE in \eqref{eqn:case2-MNE}. Later, in the simulation, we will show that the MFG converges to the MFE in \eqref{eqn:case2-MNE}, even if the condition in \eqref{eqn:convergence-case2} is not satisfied.

\section{Simulation Results and Analysis}\label{sec:simulations}

In the section, we present numerical results to illustrate  the average AoI and the average peak AoI under the two considered packet management schemes derived in Section~\ref{sec:aoi_analysis}, \textcolor{black}{the accuracy of the CTMC model}, the accuracy of the mean-field approximation, and the performance achieved by the proposed MFG framework in Section \ref{sec:mfg}.
\subsection{Illustration of the Analysis on the Average AoI and the Average Peak AoI}

\begin{figure*}[!t]
\begin{minipage}[h]{0.32\linewidth}
\centering
       \includegraphics[scale=0.35]{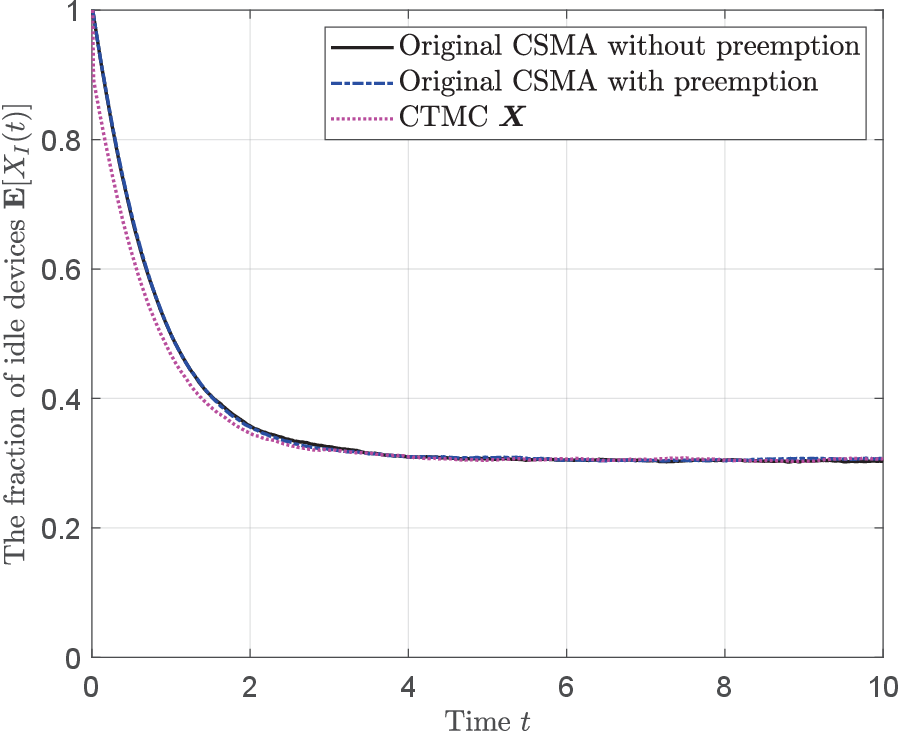}
\subcaption{}\label{fig:compare_CSMA_CTMC_X_I_N10}
\end{minipage}
\begin{minipage}[h]{0.32\linewidth}
\centering
        \includegraphics[scale=0.35]{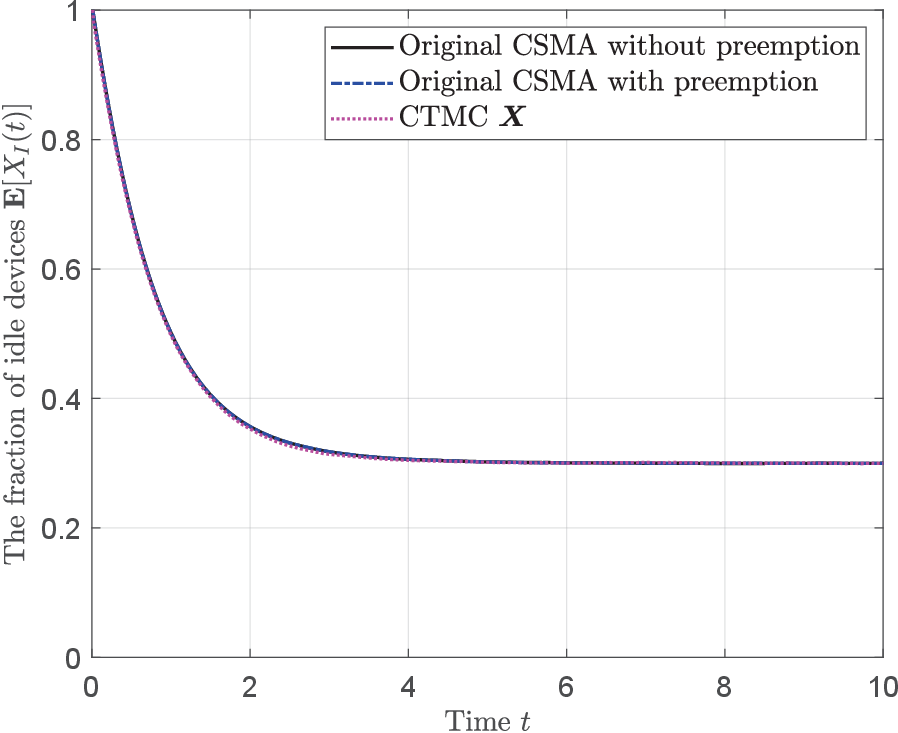}
\subcaption{}\label{fig:compare_CSMA_CTMC_X_I_N100}
\end{minipage}
\begin{minipage}[h]{0.32\linewidth}
\centering
        \includegraphics[scale=0.35]{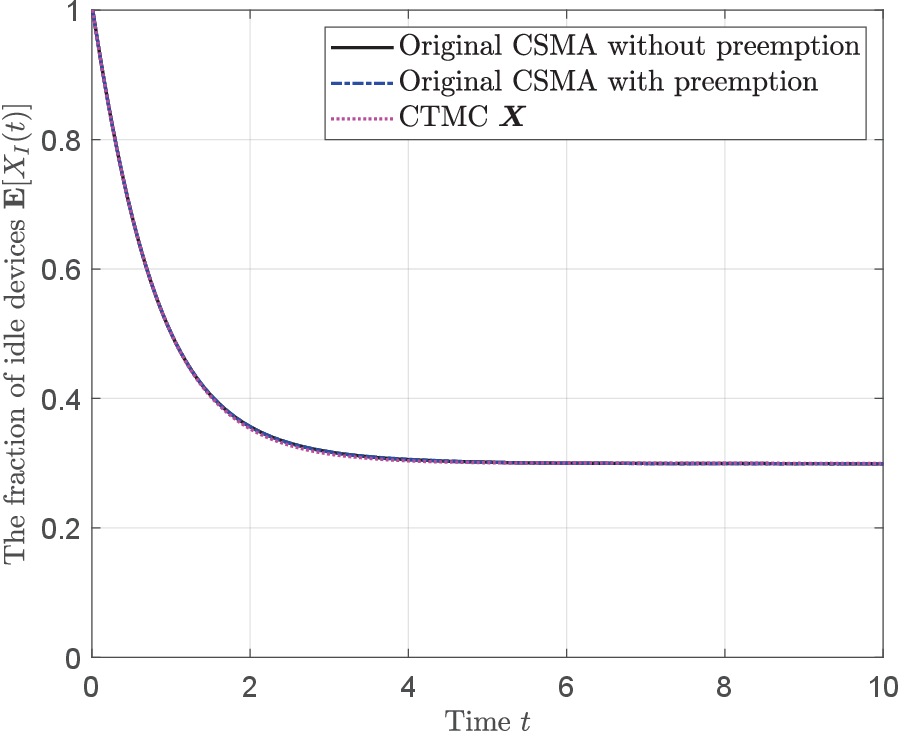}
\subcaption{}\label{fig:compare_CSMA_CTMC_X_I_N1000}
\end{minipage}

\caption{\textcolor{black}{The evolutions of the fraction of IoT devices in state (I) for the original system under CSMA with and without preemption in service, and the CTMC $\bs{X}$ in (26) under various numbers of devices $N$. $\lambda=0.8$,  $\mu=1$,  $w=2$, and $\gamma=2$.  (a) $N=10$. (b) $N=100$. (c) $N=1000$.}}
\label{fig:compare_CSMA_CTMC_X_I}

\end{figure*}

\begin{figure*}[!t]
\begin{minipage}[h]{0.32\linewidth}
\centering
       \includegraphics[scale=0.35]{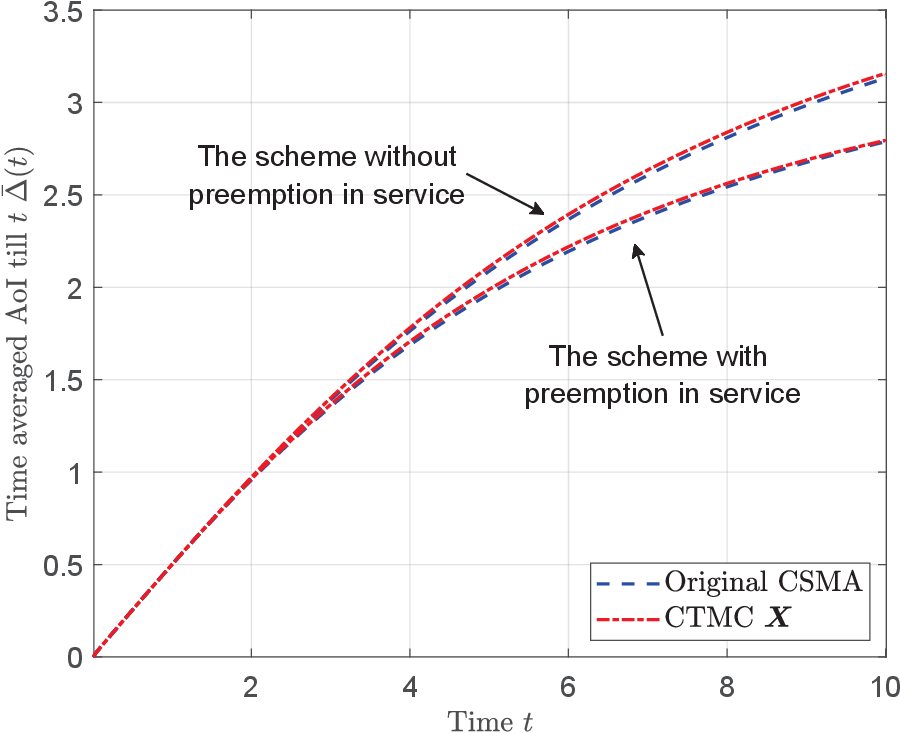}
\subcaption{}\label{fig:compare_CSMA_CTMC_AoI_N10}
\end{minipage}
\begin{minipage}[h]{0.32\linewidth}
\centering
        \includegraphics[scale=0.35]{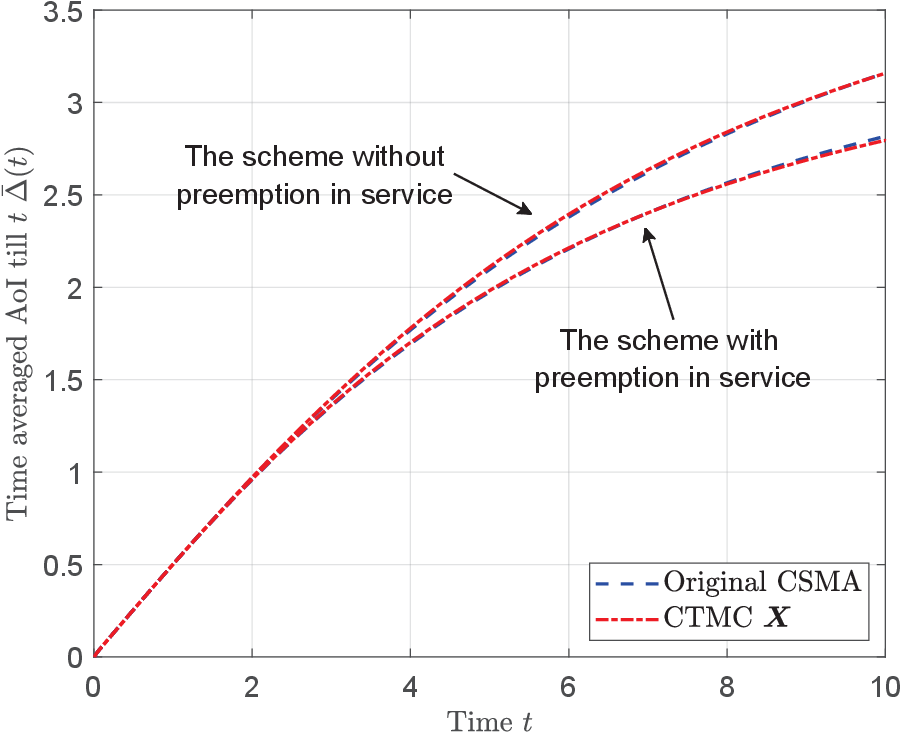}
\subcaption{}\label{fig:compare_CSMA_CTMC_AoI_N100}
\end{minipage}
\begin{minipage}[h]{0.32\linewidth}
\centering
        \includegraphics[scale=0.35]{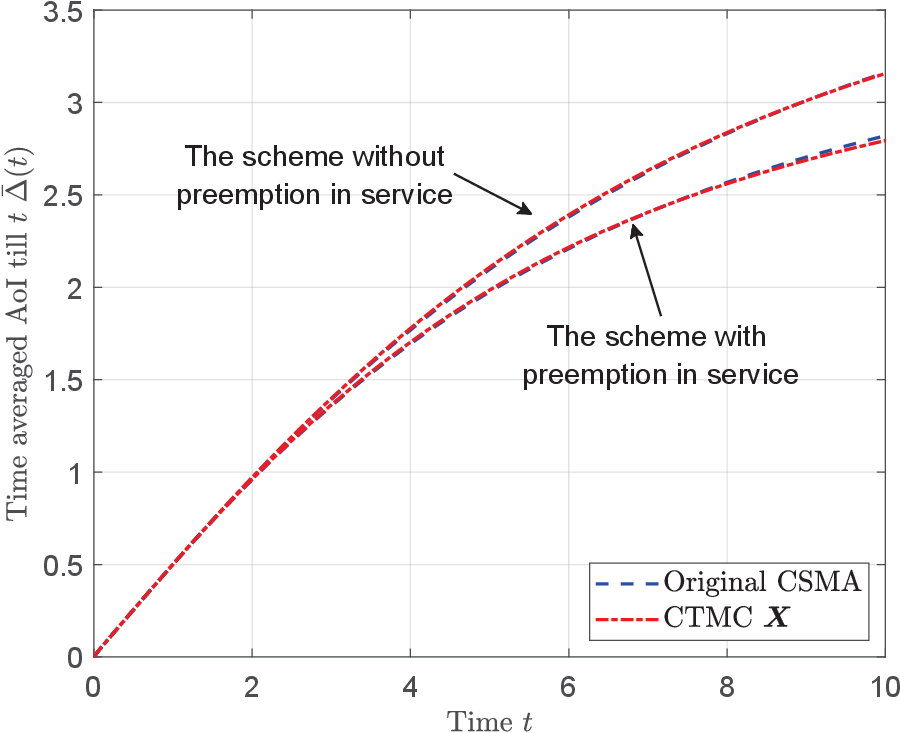}
\subcaption{}\label{fig:compare_CSMA_CTMC_AoI_N1000}
\end{minipage}

\caption{\textcolor{black}{The evolutions of of the time-averaged AoI till $t$, $\bar{\Delta}(t) = \frac{1}{t}\sum_{\tau=0}^t\Delta(\tau)$  for the original system under CSMA with and without preemption in service, and the CTMC $\bs{X}$ in (26) under various numbers of devices $N$.  $\lambda=0.8$,  $\mu=1$,  $w=2$, and $\gamma=2$.  (a) $N=10$. (b) $N=100$. (c) $N=1000$.}
}
\label{fig:compare_CSMA_CTMC_AoI}

\end{figure*}

In Fig.~\ref{fig:AoI_expression_verify}, we evaluate the analytical results in Theorem~\ref{theorem:aoi} and the simulations results of the average AoI and the average peak AoI for the packet management schemes with and without preemption in service \textcolor{black}{under a given stationary distribution}. The simulations results are obtained by averaging over 50,000 status update arrivals.
It can be seen that the simulation results agree very well with the analytical results thus corroborating the theoretical results characterized in Theorem~\ref{theorem:aoi}. Moreover, we can see that the scheme with preemption in service will always lead to better AoI performance in terms of the average AoI and the average peak AoI, than the scheme without preemption in service, the improvement in the AoI reaching up to $31\%$ and $25\%$, respectively.
We also observe that, for a fixed rate $k$, both the average AoI and the peak AoI decrease with $\lambda$.
However, this does not necessarily hold true for the case in which the rate $k$ depends on the arrival rate (in an inverse manner), as will be seen next.

\textcolor{black}
{
\subsection{Accuracy of the CTMC Model}
We consider the original system under CSMA in Section~\ref{sec:systemmodel} and the proposed CTMC $\bs{X}$ model in \eqref{eqn:transition_X}, while incorporating the AoI evolution.
The initial state of each device is set as state (I) and the simulation results are obtained by averaging over 10,000 runs of simulation trajectories.
In Fig.~\ref{fig:compare_CSMA_CTMC_X_I}, we illustrate the evolution of the fraction of IoT devices in state (I), $X_I(t)$ as a function of time for the original CSMA system with and without preemption, as well as for the CTMC $\bs{X}$ in \eqref{eqn:transition_X}. Please note that the CTMC is the same for both packet management schemes.
From Fig.~\ref{fig:compare_CSMA_CTMC_X_I}, we see that the evolution of the expected value $\mathbf{E}[X_I(t)]$ is indistinguishable for the schemes with and without preemption, and closely matches that under the CTMC, across different population sizes of $N=10$, $N=100$, and $N=1000$.
}

\textcolor{black}
{
Next, in Fig.~\ref{fig:compare_CSMA_CTMC_AoI}, we illustrate the evolution of the time-averaged AoI till time $t$, $\bar{\Delta}(t) = \frac{1}{t}\sum_{\tau=0}^t\Delta(\tau)$ as a function of time for the original system under CSMA and a system governed by the CTMC in \eqref{eqn:transition_X}.
Specifically, for the CTMC-based system with varying $N$, we consider a tagged user transmitting its status packet according to the procedures outlined in ~\ref{sec:systemmodel}, where the probability of finding an idle channel is $1-\gamma X_S(t)$. Here, $X_S(t)$ represents the fraction of devices in state (S) at $t$ obtained using the CTMC in \eqref{eqn:transition_X} with the corresponding $N$.
From ~\ref{fig:compare_CSMA_CTMC_AoI}, we can see that, the evolution of $\bar{\Delta}(t)$ under the original CSMA system and the system employing CTMC closely resemble each other, for both the scheme with and without preemption, across different population sizes of $N=10$, $N=100$, and $N=1000$.
}

\textcolor{black}
{
Note that in these simulations, we have explicitly considered the time-varying effective waiting rate $w(1-\gamma X_S(t))$ and integrated the CTMC model in \eqref{eqn:transition_X} with the AoI evolution.
Moreover, when the system reaches  the steady state $\bs{\pi}$, the effective waiting rate is nearly fixed and given by $w(1-\gamma\pi_S)$.
Hence, the simulation results in  Figs.~\ref{fig:compare_CSMA_CTMC_X_I} and \ref{fig:compare_CSMA_CTMC_AoI} confirm that the considered CTMC model in \eqref{eqn:transition_X} can accurately capture the dynamics of the original CSMA system for both packet management schemes, with and without preemption.
}

\subsection{Accuracy of the Mean-Field Approximation}
\begin{figure*}[!t]
\begin{minipage}[h]{0.32\linewidth}
\centering
       \includegraphics[scale=0.35]{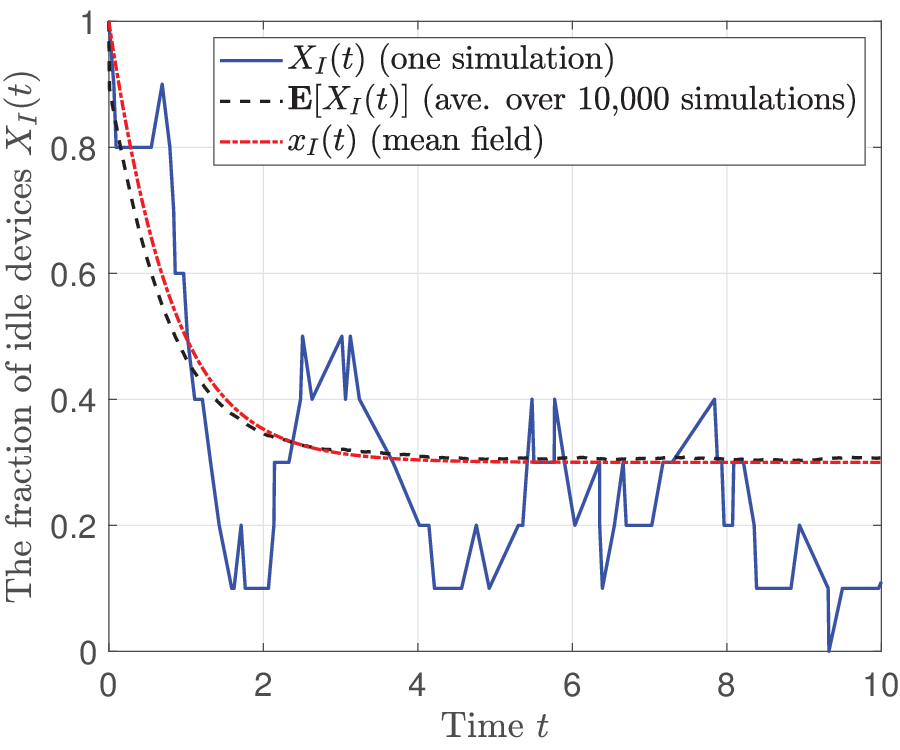}
\subcaption{}\label{fig:mean_field_N10}
\end{minipage}
\begin{minipage}[h]{0.32\linewidth}
\centering
        \includegraphics[scale=0.35]{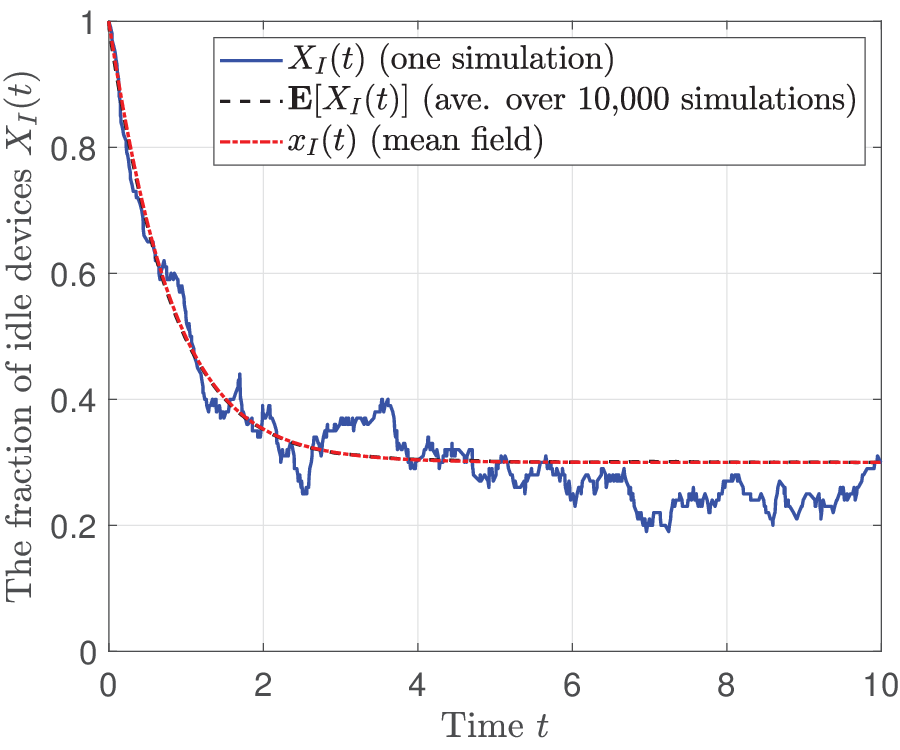}
\subcaption{}\label{fig:mean_field_N100}
\end{minipage}
\begin{minipage}[h]{0.32\linewidth}
\centering
        \includegraphics[scale=0.35]{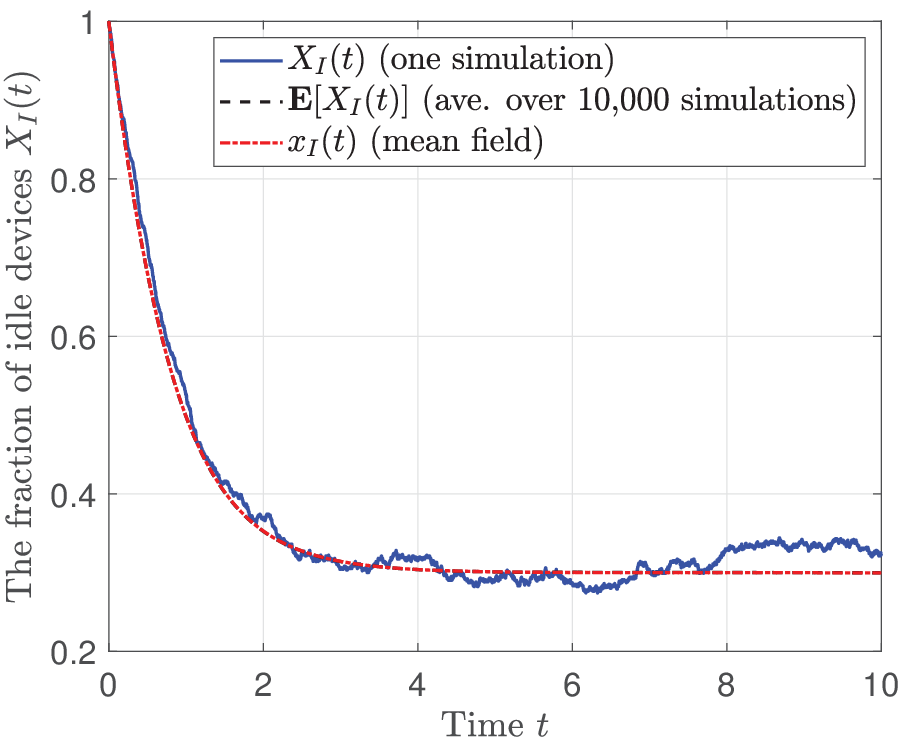}
\subcaption{}\label{fig:mean_field_N1000}
\end{minipage}

\caption{The evolutions of the fraction of IoT devices in state (I) for the CTMC $\bs{X}$ in \eqref{eqn:transition_X} under various numbers of devices $N$. The figures compare one simulation trajectory, the average of 10,000 runs of simulation trajectories and the mean-field limit obtained via the ODE in \eqref{eqn:mean-field}. $\lambda=0.8$,  $\mu=1$,  $w=2$, and $\gamma=2$.  (a) $N=10$. (b) $N=100$. (c) $N=1000$.}
\label{fig:mean_field}
\end{figure*}

Now, we consider the CMTC $\bs{X}$ in \eqref{eqn:transition_X} for different population sizes of $N=10$, $N=100$, and $N=1000$ of IoT devices and illustrate the evolution of the fraction of IoT devices in state (I), $X_I(t)$, as a function of time.
\textcolor{black}{We set the initial system state as  $\bs{X}(0)=(1,0,0)$.}
\textcolor{black}{Note that the system state dynamics with a finite $N$ can be exactly characterized by a CTMC, for which the exponential back-off times of all devices have been captured.}
 As illustrated in Fig.~\ref{fig:mean_field}, each subfigure consists of the results for one simulation trajectory, the average of 10,000 runs of simulation trajectories, and the mean-field limit obtained in the ODE in \eqref{eqn:mean-field}.
From Fig.~\ref{fig:mean_field}, we can observe that, when the number of devices $N$ increases, the simulation results of one trajectory $X_I(t)$ are concentrated on the mean-field limit $x_I(t)$.
Moreover, we observe that, even when $N=10$, the mean-field limit $x_I(t)$ is very close to the value $\mathbb{E}[X_I(t)]$ computed by averaging over 10,000 run simulations, and for $N=100$ and $N=1000$, the two curves are almost indistinguishable \textcolor{black}{(the dashed black line for $E[X_I(t)]$ hiding under the dash-dotted red line for $x_I(t)$)}.

In Table~\ref{table:steady-state}, we present the average AoI and the average peak AoI of the two packet management schemes by using the stationary distribution $\bs{\pi}$ of the CTMC $\bs{X}$ with varying number of devices and by using the equilibrium point obtained in Theorem~\ref{theorem:mean-field}. The stationary distribution $\bs{\pi}$ for each $N$ is estimated by averaging over 10,000 runs during the time period from $t=500$ to $t=1000$. It can be seen that, even for small values of $N$, the average AoI and the average peak AoI are very close to the \textcolor{black}{values} obtained in the mean-field limit. Hence, Fig.~\ref{fig:mean_field} and Table~\ref{table:steady-state} indicate that our proposed mean-field approximation is very accurate for the considered system.

\begin{table*}[!t]
\centering
\caption{Average AoI and average peak AoI under the stationary distribution $\bs{\pi}$ obtained via simulations and the equilibrium point $\bs{x}^*$ of the ODE in \eqref{eqn:mean-field}.}
 \label{table:steady-state}
\resizebox{0.95\textwidth}{!}{%
\begin{tabular}{@{}lcccccc@{}}
\toprule
AoI                                 & $N=10$     & $N=20$     & $N=50$     & $N=100$    & $N=1000$  & Mean-field \\ \midrule
Avg. AoI with preemption         & 3.820702 & 3.820489 & 3.820589 & 3.820453 & 3.82068 & 3.811444   \\
Avg. peak AoI with preemption    & 5.159022 & 5.158755 & 5.15888  & 5.158710  & 5.15900   & 5.147431   \\
Avg. AoI without preemption      & 4.602450  & 4.602219 & 4.602327 & 4.602181 & 4.60243 & 4.592457   \\
Avg. peak AoI without preemption & 5.940769 & 5.940485 & 5.940618 & 5.940438 & 5.94074 & 5.928443   \\ \bottomrule
\end{tabular}%
}
\end{table*}

\begin{figure}[!ht]
\begin{centering}
\includegraphics[scale=.45]{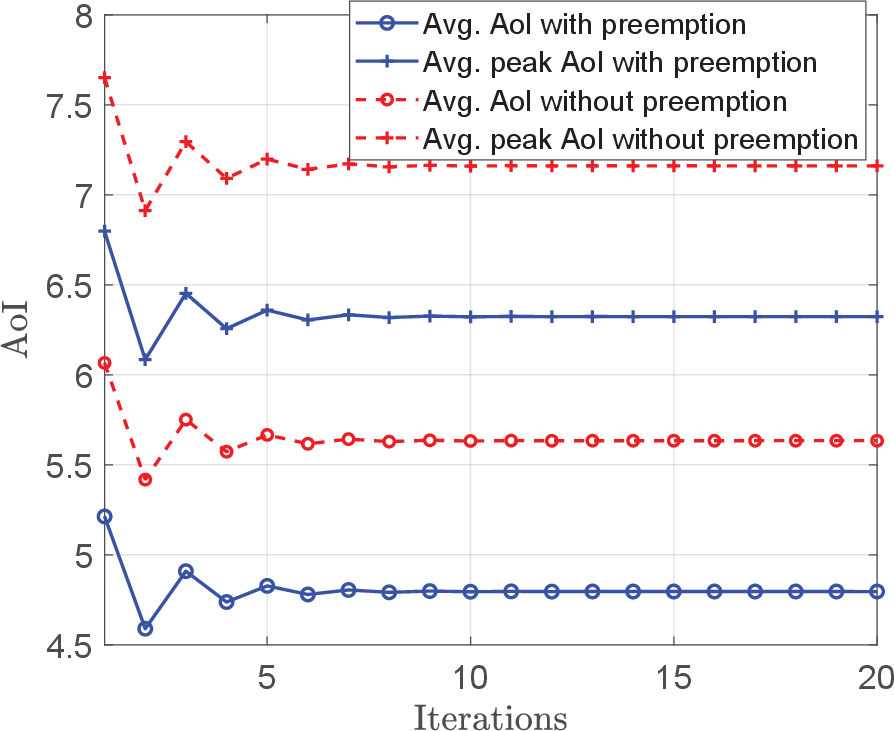}
 \caption{Illustration of the convergence of the MFG. $\lambda=0.8$, $\mu=1$, and $\gamma=5$.}\label{fig:convergence_verify}
\end{centering}
\end{figure}

\subsection{Performance of the Proposed MFG}
Now, we illustrate the AoI performance achieved by our proposed MFG. In this subsection, we set the channel sensing cost per unit time to $C_s=0.1$, the transmission cost per unit time to $C_t=0.2$, and the average energy constraint to $\hat{C}=0.4$.

In Fig.~\ref{fig:convergence_verify}, we show the evolution of the average AoI and the average peak AoI resulting from the proposed MFG for the two packet management schemes. We set $\lambda=0.8$, $\mu=1$, and $\gamma=5 $. In this case, we can verify that the condition  in \eqref{eqn:case2-MNE}  for Case 2) is satisfied.
From Fig.~\ref{fig:convergence_verify}, we  observe that the proposed MFG converges quickly to the MFE in \eqref{eqn:case2-MNE} in Case 2), although the sufficient condition in \eqref{eqn:convergence-case2} for the convergence to the MFE in Case 2) is not satisfied.
This indicates the proposed MFG possesses good convergence properties.

\begin{figure*}[!t]
\begin{minipage}[h]{0.32\linewidth}
\centering
       \includegraphics[scale=0.35]{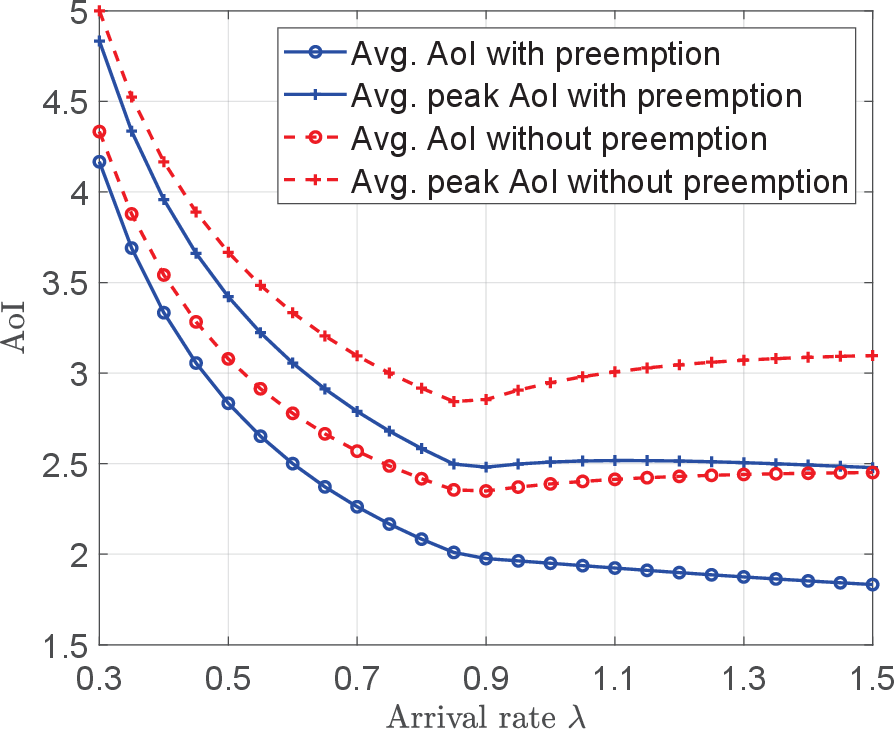}
\subcaption{}\label{fig:opt_vs_lambda_gamma2}
\end{minipage}
\begin{minipage}[h]{0.32\linewidth}
\centering
        \includegraphics[scale=0.35]{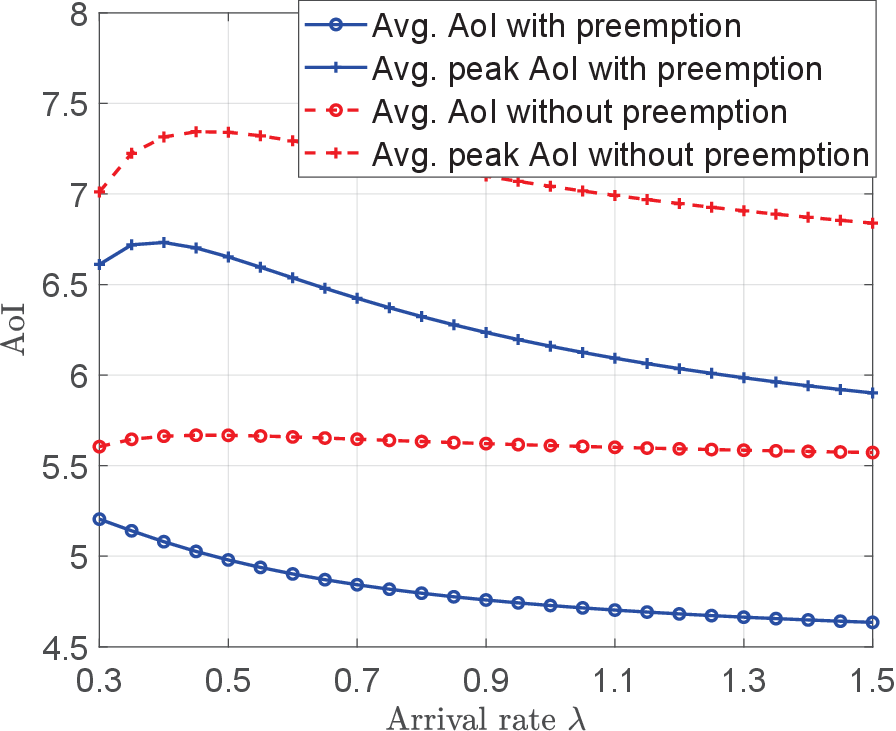}
\subcaption{}\label{fig:opt_vs_lambda_gamma5}
\end{minipage}
\begin{minipage}[h]{0.32\linewidth}
\centering
        \includegraphics[scale=0.35]{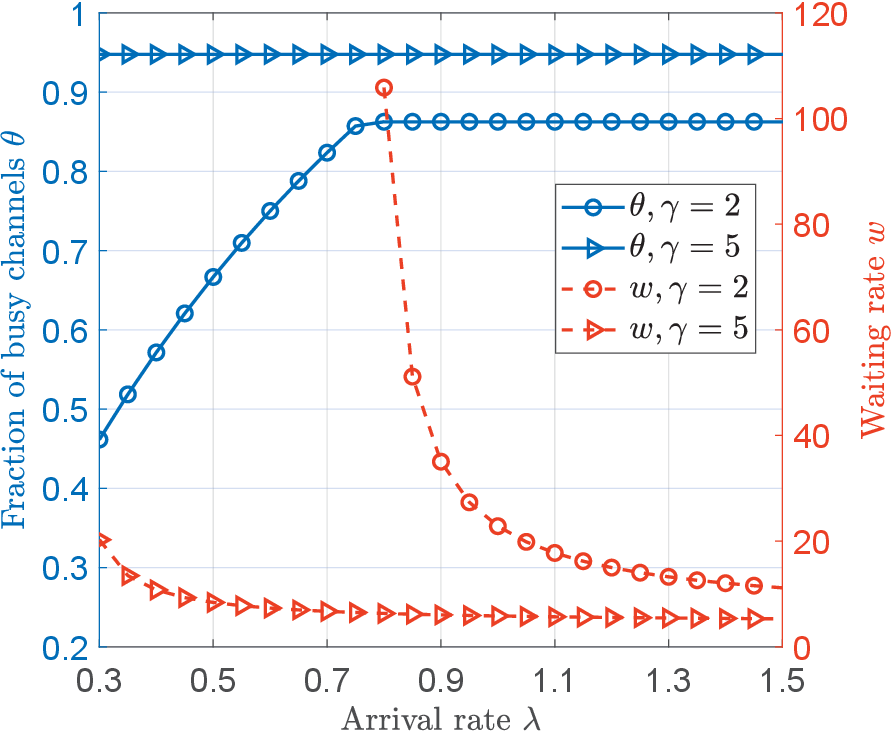}
\subcaption{}\label{fig:opt_vs_lambda_theta_wgamma25}
\end{minipage}
\caption{Performance achieved by the MFG under varying arrival rates $\lambda$. $\mu=1$. (a) The average AoI and the average peak AoI for $\gamma=2$. (b) The average AoI and the average peak AoI for $\gamma=5$. (c) The fraction of busy channels $\theta$ and the waiting rate for $\gamma=2,5$.}\label{fig:opt_lambda}
\end{figure*}

\begin{figure*}[!t]
\begin{minipage}[h]{0.32\linewidth}
\centering
       \includegraphics[scale=0.35]{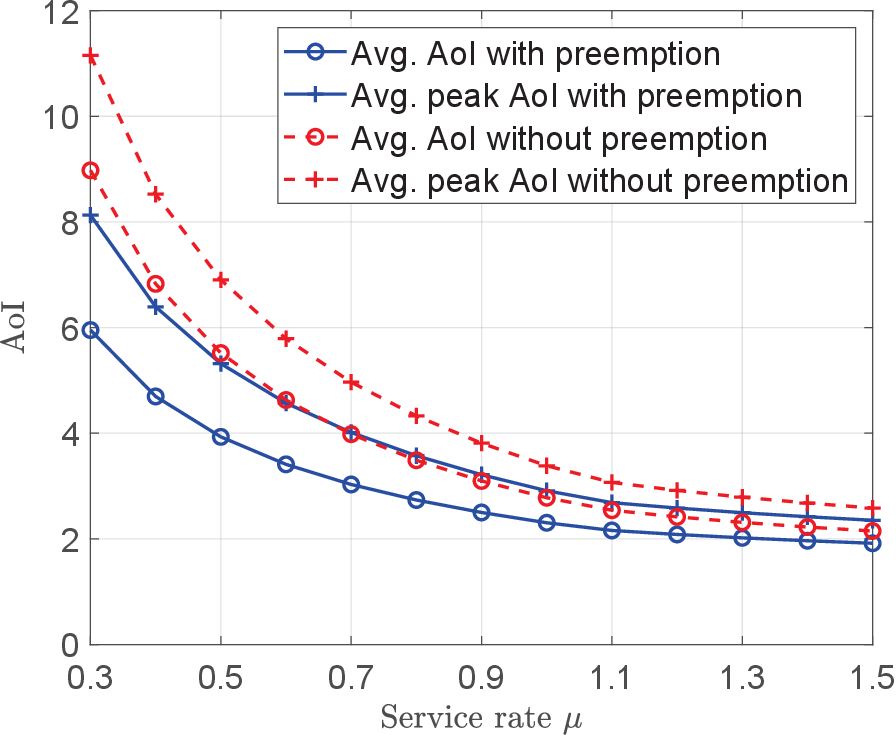}
\subcaption{}\label{fig:opt_vs_mu_gamma2}
\end{minipage}
\begin{minipage}[h]{0.32\linewidth}
\centering
        \includegraphics[scale=0.35]{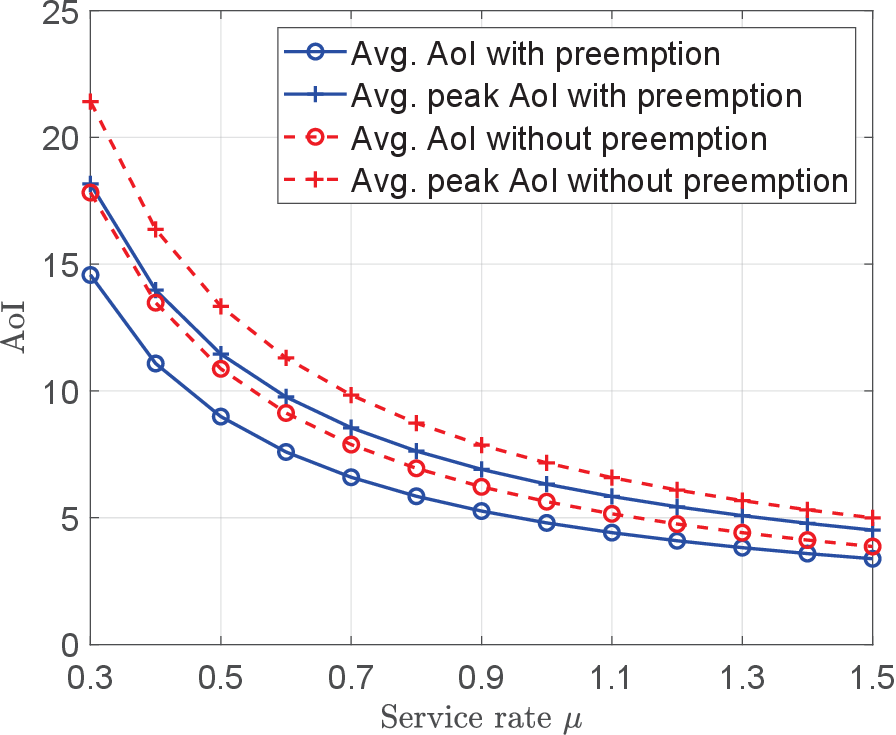}
\subcaption{}\label{fig:opt_vs_mu_gamma5}
\end{minipage}
\begin{minipage}[h]{0.32\linewidth}
\centering
        \includegraphics[scale=0.35]{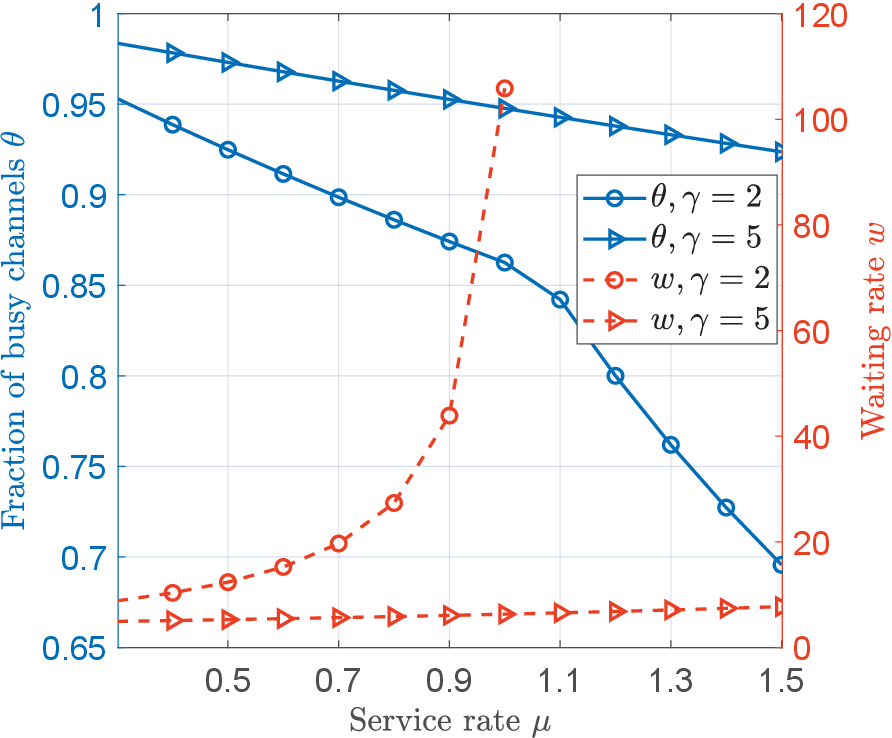}
\subcaption{}\label{fig:opt_vs_mu_theta_wgamma25}
\end{minipage}
\caption{Performance achieved by the MFG under varying service rates $\mu$. $\lambda=0.8$. (a) The average AoI and the average peak AoI for $\gamma=2$. (b) The average AoI and the average peak AoI for $\gamma=5$. (c) The fraction of busy channels $\theta$ and the waiting rate for $\gamma=2$ and $\gamma=5$.}\label{fig:opt_mu}
\end{figure*}

Fig.~\ref{fig:opt_lambda} illustrates the performance of the MFE achieved by the proposed MFG under different  arrival rates. In particular, in Fig.~\ref{fig:opt_vs_lambda_gamma2} and Fig.~\ref{fig:opt_vs_lambda_gamma5}, we show the average AoI and the average peak AoI under the two packet management schemes for $\gamma=2$ and $\gamma=5$, respectively, and in Fig.~\ref{fig:opt_vs_lambda_theta_wgamma25}, we illustrate the fraction of busy channels $\theta$ and the waiting rate $w$ for  $\gamma=2$ and $\gamma=5$.
From Fig.~\ref{fig:opt_vs_lambda_theta_wgamma25}, we observe that for $\gamma=2$,  $\lambda\in[0.3,0.75]$ corresponds to the MFE in Case 1) and $\lambda\in [0.8,1.5]$ corresponds to the MFE in Case 2); and for $\gamma=2$, $\lambda\in[0.3,1.5]$ corresponds to the MFE in Case 2).
Then, from Fig.~\ref{fig:opt_vs_mu_gamma2}, we can see that the average AoI and the average peak AoI always decrease with the arrival rate $\lambda$ in Case 1). This can be justified by the closed-form expressions for $k\to\infty$ derived in \eqref{eqn:avg_aoi_wop_limit} and \eqref{eqn:avg_aoi_wp_limit}.
However, from Fig.~\ref{fig:opt_vs_lambda_gamma2} and Fig.~\ref{fig:opt_vs_lambda_gamma5}, in Case 2), we can see that, the average AoI and the average peak AoI do not necessarily decrease with $\lambda$. This is because, in Case 2), as $\lambda$ increases, the waiting rate $w$ decreases so as to keep the fraction of busy channels $\theta$ fixed, as seen from Fig.~\ref{fig:opt_vs_lambda_theta_wgamma25}.
Moreover, from Fig.~\ref{fig:opt_vs_lambda_theta_wgamma25}, we can see that, the waiting rate $w$ decreases with $\gamma$. This implies that with fewer communication resources (i.e., a smaller $\gamma=N/M$), each device has to wait for a longer time period (i.e., a larger mean $1/w$) before transmitting its status update.

Similarly, in Fig.~\ref{fig:opt_mu}, we illustrate the average AoI and the average peak AoI of the two schemes, the fraction of busy channels $\theta$, and the waiting rate $w$ in the MFE, under different  service rates. From Fig.~\ref{fig:opt_vs_mu_theta_wgamma25}, we can see that for $\gamma=2$, $\mu\in[0.3,1.0]$ corresponds to the MFE in Case 2) and $\mu\in[1.1,1.9]$ corresponds to the MFE in Case 1); and for $\gamma=5$, $\mu\in[0.3,1.9]$ corresponds to the MFE in Case 2). Then, from Fig.~\ref{fig:opt_vs_mu_gamma2} and Fig.~\ref{fig:opt_vs_mu_gamma5}, we observe that the average AoI and the average peak AoI always decrease with the service rate $\mu$ in both Case 1) and Case 2). This is due to the fact, for a larger $\mu$, the rate $k=w(1-\theta)$ increases, as the waiting rate $w$ increases and the fraction of busy channels $\theta$ decreases, as seen from Fig.~\ref{fig:opt_vs_mu_theta_wgamma25}.
This confirms the intuitive observation that channels with better conditions  lead to smaller AoI.

\begin{figure}[!t]
\begin{centering}
\includegraphics[scale=.45]{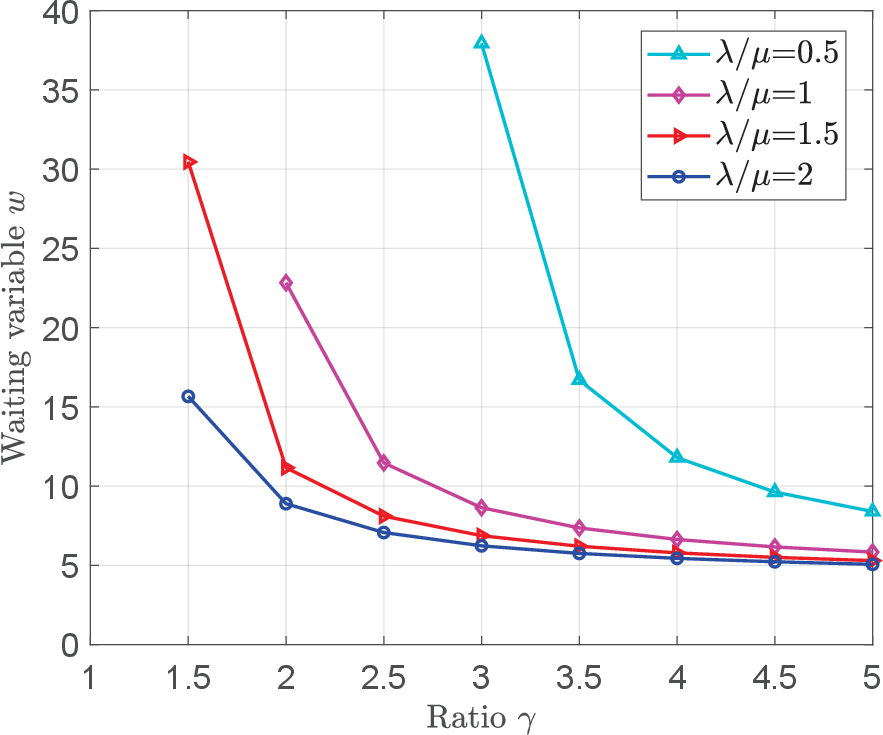}
 \caption{The waiting rate achieved by the MFG under varying ratio $\gamma$ and channel utilization load $\lambda/\mu$. $\mu=1$.}\label{fig:opt_gamma}
\end{centering}
\end{figure}

    In Fig.~\ref{fig:opt_gamma}, we illustrate the waiting rate $w$ in the MFE under different values for the ratio $\gamma\in\{1,1.5,\cdots,5\}$ and the channel utilization load $\lambda/\mu$. Here, we do not plot the points of $w=\infty$ for the MFE in Case 1) and only plot the points of $w$ for the MFE in Case 2).
For example, it can be seen that for $\lambda/\mu= 0.5$, $\gamma\in\{1,1.5,2,2.5\}$ corresponds the MFE in Case 1 and $\gamma\in\{3,3.5,4,4.5,5\}$ corresponds the MFE in Case 2.
From Fig.~\ref{fig:opt_gamma}, we can see that, for a given $\lambda/\mu$, under the MFE in Case 2), the waiting rate decreases with the ratio $\gamma$.
The reason is that, with fewer communication resources, each device has to wait for a longer period before transmitting its packet.
Moreover, we observe that, for a given $\gamma$, with the decrease of $\lambda/\mu$, the MFE is more likely to be in Case 1).
This indicates that, if the channel utilization load is lighter, each device is more likely to transmit without waiting.

\begin{figure}[!t]
\begin{minipage}[h]{1\linewidth}
\centering
       \includegraphics[scale=0.45]{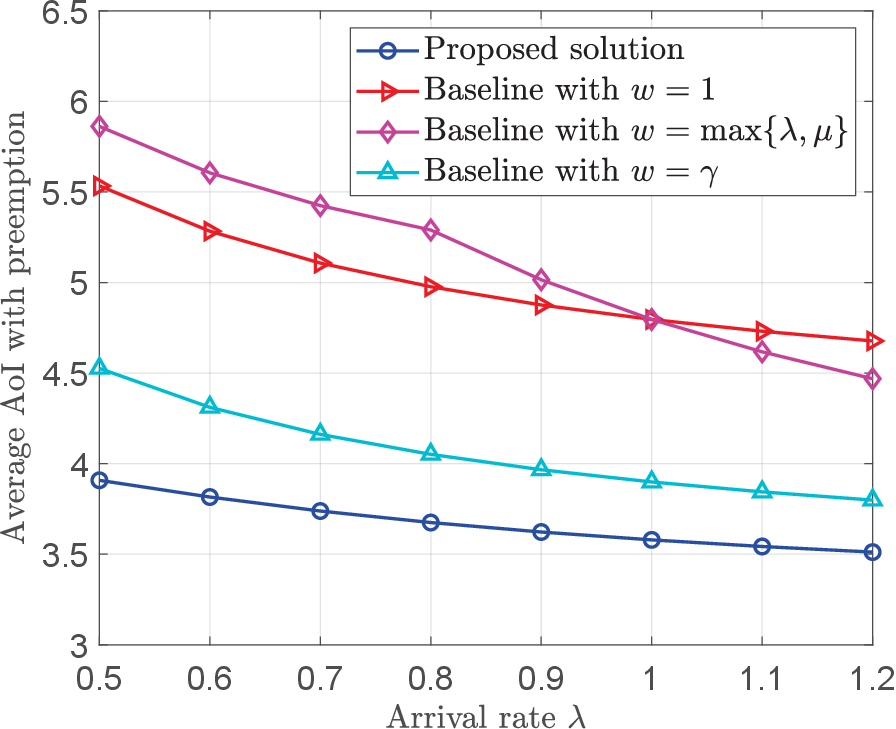}
\subcaption{}\label{fig:perf_vs_lambda_gamma5}
\end{minipage}\\
\begin{minipage}[h]{1\linewidth}
\centering
        \includegraphics[scale=0.45]{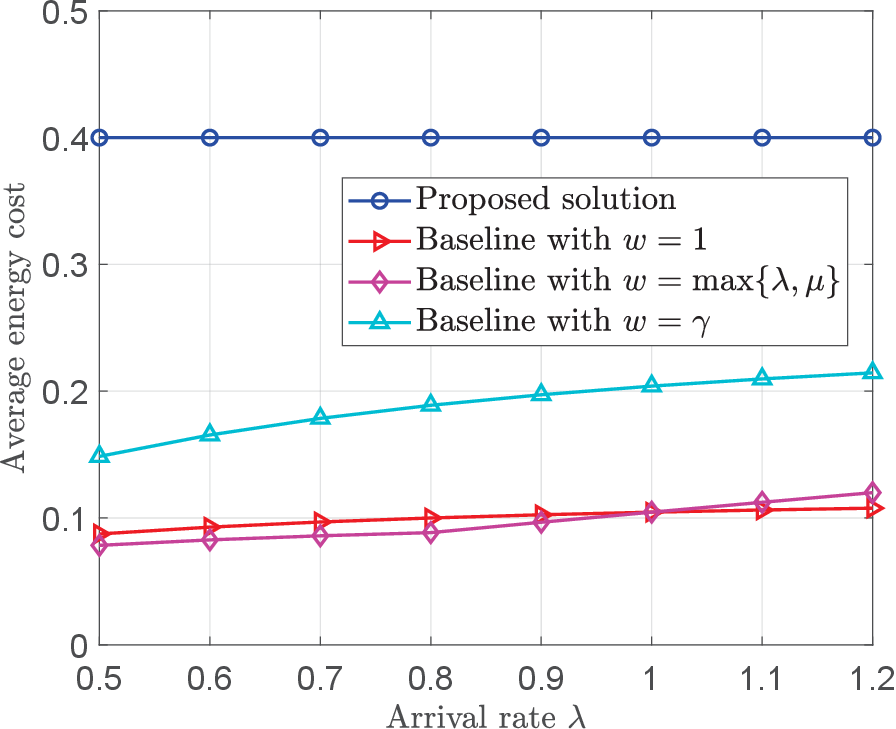}
\subcaption{}\label{fig:perf_energy_vs_lambda_gamma5}
\end{minipage}
\caption{Performance comparison among the proposed MFG, the baseline with $w=1$, the baseline with $w=\max\{\lambda,\mu\}$, and the baseline with $w=\gamma$ under varying arrival rates $\lambda$. $\mu=1$ and $\gamma=3$. (a) The average AoI under the scheme with preemption in service. (b) The average energy cost.}\label{fig:perf_lambda}
\end{figure}

\begin{figure}[!t]
\begin{minipage}[h]{1\linewidth}
\centering
       \includegraphics[scale=0.45]{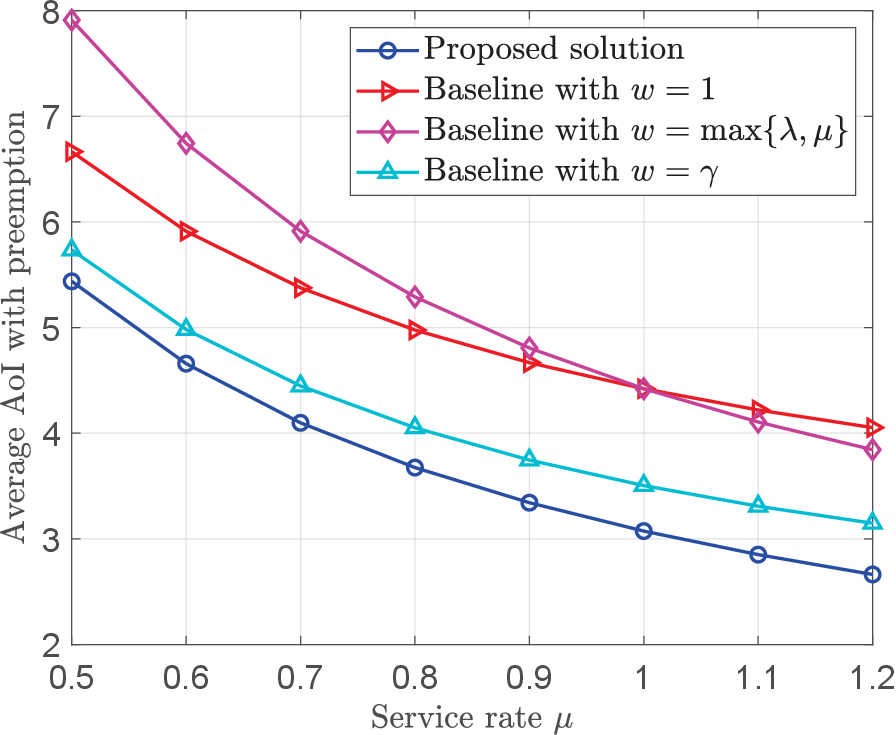}
\subcaption{}\label{fig:perf_vs_mu_gamma5}
\end{minipage}

\begin{minipage}[h]{1\linewidth}
\centering
        \includegraphics[scale=0.45]{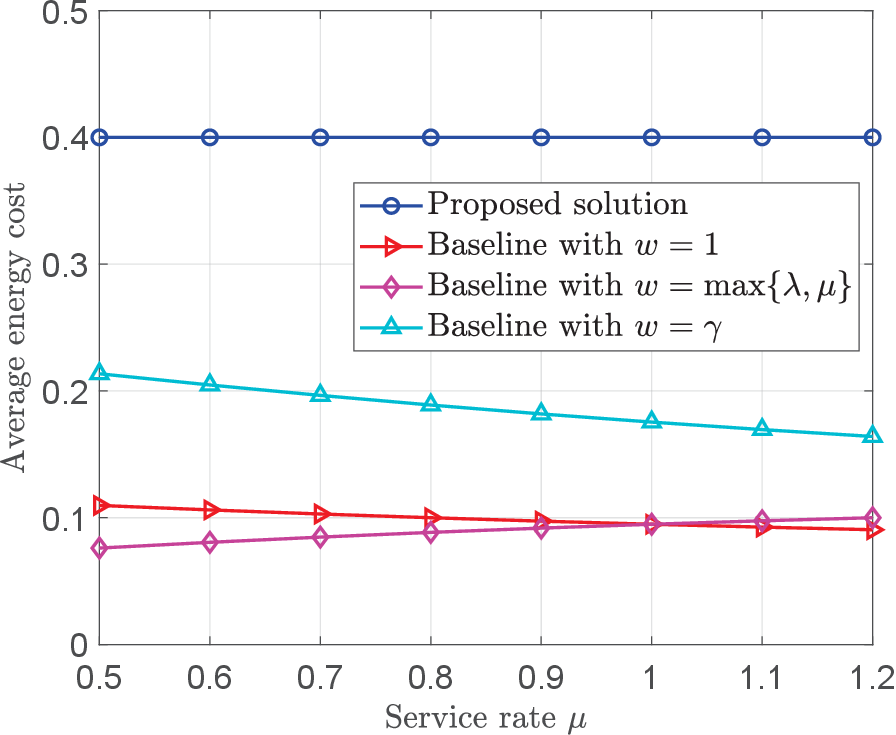}
\subcaption{}\label{fig:perf_energy_vs_mu_gamma5}
\end{minipage}
\caption{Performance comparison among the proposed MFG, the baseline with $w=1$, the baseline with $w=\max\{\lambda,\mu\}$, and the baseline with $w=\gamma$, under varying arrival rates $\mu$. $\lambda=0.8$ and $\gamma=3$. (a) The average AoI under the scheme with preemption in service. (b) The average energy cost.}\label{fig:perf_mu}
\end{figure}

\textcolor{black}{
Finally, in Fig.~\ref{fig:perf_lambda} and Fig.~\ref{fig:perf_mu}, we compare the average AoI under the scheme with preemption and the average energy cost, resulting from our proposed MFG, a baseline policy with $w=1$, a baseline policy with $w=\max\{\lambda,\mu\}$, and a baseline policy with $w=\gamma$ under varying $\lambda$ and $\mu$, respectively.
In Fig.~\ref{fig:perf_vs_lambda_gamma5} and Fig.~\ref{fig:perf_vs_mu_gamma5}, we observe that, the average AoI of the proposed MFG and the three baseline polices decrease with $\lambda$ and $\mu$.
Clearly, the average AoI improvement achieved by the proposed MFG compared with the baselines with $w=1$, $w=\max\{\lambda,\mu\}$, and $w=\gamma$ can reach up to to 28\%, 32\%, and 12\% respectively.
This is due to the fact,  as seen from Fig.~\ref{fig:perf_energy_vs_lambda_gamma5} and Fig.~\ref{fig:perf_energy_vs_mu_gamma5}, the proposed MFG can make better use of the available energy at the IoT device, than the three baselines. This demonstrates the effectiveness of the proposed MFG. Here, we would like to mention that, the considered three baseline policies may not guarantee that the average energy cost constraint is always satisfied. On the contrary, under our proposed MFG, the average energy constraint is always satisfied.}

\section{Conclusion}\label{sec:conclusion}
In this paper, we have investigated a CSMA-type random access scheme for an ultra-dense IoT monitoring system, under which multiple devices contend for channel access and communicate their status updates to the associated receivers.
We have derived the closed-form expressions of the average AoI and the average peak AoI for two packet management schemes with and without preemption in service.
We have shown that the scheme with preemption in service can always lead to smaller average AoI and smaller average peak AoI than the scheme without preemption in service.
Then, we have formulated a noncooperative game in which each device aims at optimizing its waiting rate so as to minimize its average AoI or its average peak AoI in the two packet management schemes, under an energy cost constraint.
To overcome the difficulties in solving this problem for a dense IoT with a large number of devices, we have proposed an MFG framework to study the asymptotic performance of each device when the number of devices goes to infinity.
We have shown that the MFG approximation is suitable for the considered system, and, then, we have conducted a comprehensive analysis of the existence, the uniqueness, and the convergence properties of the MFE.
We have shown that each device can transmit without waiting if there are sufficient communication resources and the channel utilization load is low.
Simulation results validate the correctness of the derived closed-form expressions of the average AoI and the average peak AoI. These results show that the proposed MFG is very accurate, even for very small numbers of devices and demonstrate the effectiveness of the proposed CSMA-type scheme under the MFG over three baseline schemes. Future works will address key extensions such as studying a heterogeneous IoT monitoring system and investigating the contention resolution problem in CSMA with non-zero sensing delay\cite{9007478,bedewy2019optimizing,5340575}.
It would also be interesting to consider more general distributions for the transmission time \cite{najm2016age}.

\appendices

\section{Proof of Theorem~\ref{theorem:mean-field}}\label{app:mean-field}
First, we show that the equilibrium point  $\bs{x}^*=(x_I^*,x_W^*,x_S^*)$ of the mean-field model in \eqref{eqn:mean-field} is unique and satisfies \eqref{eqn:mean-field-solution}.
According to  \eqref{eqn:mean-field}, and by the definition that $x_I^*+x_W^*+x_S^*=1$, we can see that the equilibrium point $\bs{x}^*$ satisfies the following fixed point equations:
\begin{subequations}\label{eqn:mean-field-solution-proof}
\begin{align}
x_I^* &= \frac{\mu}{\lambda}x_s^*,\\
x_W^* &= \frac{\mu x_S^*}{w(1-\gamma x_S^*)},\\
x_S^* &= \frac{1/\mu}{1/\lambda+1/(w_n(1-\gamma x_S^*))+1/\mu} \nonumber\\
&= \frac{\lambda w (1-\gamma x_S^*)}{(\lambda+\mu)w (1-\gamma x_S^*) + \lambda\mu}.\label{eqn:xs}
       \end{align}
\end{subequations}
By \eqref{eqn:mean-field-solution-proof}, we only need to show the uniqueness of $x_S^*$ in \eqref{eqn:xs}.
We transform \eqref{eqn:xs} into the following quadratic equation:
\begin{align}\label{eqn:solving_x_s}
w(\lambda+\mu)\gamma (x_S^*)^2 - (w(\lambda+\mu+\lambda\gamma)+\lambda\mu)x_S^* + \lambda w=0.
\end{align}
for which, the two possible solutions are given by:
\begin{align}
x_{S,1}^*=&\frac{1}{2w\gamma(\lambda+\mu)}\Big(w(\lambda + \mu + \lambda \gamma) + \lambda\mu \nonumber\\
&- \sqrt{(w(\lambda + \mu + \lambda \gamma) + \lambda\mu)^2 -4\lambda (\lambda + \mu)\gamma w^2}\Big),\label{eqn:xs1}\\
x_{S,2}^*=&\frac{1}{2w\gamma(\lambda+\mu)}\Big(w(\lambda + \mu + \lambda \gamma) + \lambda\mu \nonumber\\
&+ \sqrt{(w(\lambda + \mu + \lambda \gamma) + \lambda\mu)^2 -4\lambda (\lambda + \mu)\gamma w^2}\Big).\label{eqn:xs2}
\end{align}
Now, we show that only $x_{S,1}^*$ is feasible.
We begin by proving
\begin{align}\label{eqn:proof-1}
&(w(\lambda + \mu + \lambda \gamma) + \lambda\mu)^2 -4\lambda (\lambda + \mu)\gamma w^2\nonumber\\
&\hspace{20mm}> (w(\lambda + \mu + \lambda \gamma) + \lambda\mu - 2w(\lambda+\mu))^2
\end{align}
holds. After some algebraic manipulation, we show that \eqref{eqn:proof-1} is equivalent to $\lambda\mu>0$, and thus \eqref{eqn:proof-1}  holds.
Then, we note that the probability that a channel is sensed busy is $\gamma x_S^*$, which can be seen in Fig.~\ref{fig:ctmc}.
Thus, we must have $0\leq x_S^*\leq \frac{1}{\gamma}$. By substituting \eqref{eqn:proof-1} into \eqref{eqn:xs1} and \eqref{eqn:xs2}, respectively, we can obtain:
\begin{align}
&0<x_{S,1}^*<\frac{1}{2w\gamma(\lambda+\mu)}\Big(w(\lambda + \mu + \lambda \gamma) + \lambda\mu\nonumber\\
&\hspace{10mm}- (w(\lambda + \mu + \lambda \gamma) + \lambda\mu - 2w(\lambda+\mu))\Big)=\frac{1}{\gamma},\\
&x_{S,2}^*>\frac{1}{2w\gamma(\lambda+\mu)}\Big(w(\lambda + \mu + \lambda \gamma) + \lambda\mu \nonumber\\
&-w(\lambda + \mu + \lambda \gamma) - \lambda\mu + 2w(\lambda+\mu)\Big)=\frac{1}{\gamma}.
\end{align}
Therefore, we have shown that only $x_{S,1}^*$ is feasible, which completes the proof for the existence and the uniqueness of the equilibrium point of the mean-field model in \eqref{eqn:mean-field}.

Next, we prove the convergence to the equilibrium point in \eqref{eqn:mean-field-solution} by applying  \cite[Theorem 3.2]{10.1145/3084454}. In particular, we need to show that $\bs{x}^*$ is locally exponentially stable and is globally asymptotically stable (i.e., a unique attractor to which all trajectories converge).
For the nonlinear dynamic system in \eqref{eqn:mean-field}, we first introduce the corresponding linearized  system at its equilibrium point $\bs{x}^*$.
By deriving the Jacobian matrix of \eqref{eqn:mean-field}:
\begin{align}
\frac{\partial f}{\partial \bs{x}} =  \left( \begin{matrix} -\lambda & 0& \mu \\
\lambda & -w(1-\gamma x_S) & w\gamma x_W \\
0 &w(1-\gamma x_S) & -w\gamma x_W -\mu \end{matrix} \right)
\end{align}
we can obtain the linearized system at the equilibrium point $\bs{x}^*$, given by
\begin{subequations}\label{eqn:linearized}
\begin{align}
        &\dot{q}_I = -\lambda q_I + \mu q_S,\\
        &\dot{q}_W =  \lambda q_I -w(1-\gamma x_S^*) q_W + w\gamma x_W^* q_S,\\
        &\dot{q}_S = w(1-\gamma x_S^*) q_W - (w\gamma x_W^*+\mu) q_S ,
\end{align}
\end{subequations}
where $\bs{q} \triangleq (q_I, q_W, q_S) = \bs{x}- \bs{x}^*$.

According to  \cite[Theorem 4.15]{Khalil2002}, $\bs{x}=\bs{x}^*$ is an  exponentially stable equilibrium point for the nonlinear system in \eqref{eqn:mean-field} if and only if the corresponding linearized system at point $\bs{x}^*$ in \eqref{eqn:linearized} is exponentially stable.
We use the Lyapunov method to prove that the linearized system in \eqref{eqn:linearized} is exponentially stable.
Define the Lyapunov function as $V(t) = |q_I| + |q_W| + |q_S|$.
 Note that $q_I + q_W + q_S =0$.
 Thus, if $\bs{q}\neq 0$, then $\bs{q}$  contains either  only one positive element or two positive elements (i.e., only one negative element).
 \begin{enumerate}
\item If $q_I$ is the only one positive element, i.e.,   $q_I>0$, $q_W\leq 0$, and $q_S\leq 0$, then we have
 \begin{align}
    &V(t) = q_I - q_W - q_S = 2q_I,\\
    &\dot{V}(t) =2\dot{q}_I = -2\lambda q_I +2 \mu q_S \leq -2\lambda q_I,
 \end{align}
otherwise, if $q_I$ is the only one negative element, i.e., $q_I<0$, $q_W> 0$, and $q_S> 0$, then,
 \begin{align}
&V(t) = -q_I + q_W + q_S = -2q_I,\\
&\dot{V}(t) =-2\dot{q}_I = 2\lambda q_I -2 \mu q_S \leq 2\lambda q_I.
 \end{align}
 Thus, we have $\dot{V}(t)\leq -\lambda V(t)$ in these two cases.

\item  If $q_W$ is the only one positive element, i.e.,   $q_W>0$, $q_I\leq 0$, and $q_S\leq 0$, then we have
 \begin{align}
    &V(t) = -q_I + q_W - q_S = 2q_W,\\
    &\dot{V}(t) =2\dot{q}_W = 2\lambda q_I -2w(1-\gamma x_S^*) q_W + 2w\gamma x_W^* q_S \nonumber\\
  &\hspace{18mm}\leq -2w(1-\gamma x_S^*) q_W,
 \end{align}
otherwise, if $q_W$ is the only one negative element, i.e., $q_W<0$, $q_I> 0$, and $q_S> 0$, then,
 \begin{align}
&V(t) = q_I - q_W + q_S = -2q_W,\\
&\dot{V}(t) =2\dot{q}_W = -2\lambda q_I +2w(1-\gamma x_S^*) q_W - 2w\gamma x_W^* q_S \nonumber\\
  &\hspace{18mm}\leq 2w(1-\gamma x_S^*) q_W.
 \end{align}
 Thus, we have $\dot{V}(t)\leq -w(1-\gamma x_S^*) V(t)$ in these two cases.

\item If $q_S$ is the only one positive element, i.e.,   $q_S>0$, $q_I\leq 0$, and $q_W\leq 0$, then we have
 \begin{align}
    &V(t) = -q_I - q_W + q_S = 2q_S,\\
    &\dot{V}(t) =2w(1-\gamma x_S^*) q_W - 2(w\gamma x_W^*+\mu) q_S \nonumber\\
  &\hspace{18mm}\leq - 2(w\gamma x_W^*+\mu) q_S,
 \end{align}
otherwise, if $q_S$ is the only one negative element, i.e., $q_S<0$, $q_I> 0$, and $q_W> 0$, then,
 \begin{align}
&V(t) = -q_I + q_W - q_S = -2q_S,\\
&\dot{V}(t) =2\dot{q}_W = -2w(1-\gamma x_S^*) q_W + 2(w\gamma x_W^*+\mu) q_S \nonumber\\
  &\hspace{18mm}\leq 2(w\gamma x_W^*+\mu) q_S.
 \end{align}
 Thus, we have $\dot{V}(t)\leq -(w\gamma x_W^*+\mu) V(t)$ in these two cases.
\end{enumerate}

It can be easily checked that the above inequalities hold for $\bs{q}=0$.
Therefore, we have
\begin{align}\label{eqn:proof_thoerem1}
\dot{V}(t)\leq -\delta V(t),
\end{align}
where $\delta = \min\{\lambda,w(1-\gamma x_S^*),w\gamma x_W^* \}>0$,
which implies that
\begin{align}
|q_I| + |q_W| + |q_S| = V(t) \leq V(0)e^{-\delta t}.
\end{align}
Thus, the linearized system in \eqref{eqn:linearized} is exponentially stable, implying that the equilibrium point $\bs{x}^*$ of the mean-field model in \eqref{eqn:mean-field} is (locally) exponentially stable.
Moreover, by the Lyapunov theorem in \cite[Theorem 4.1]{Khalil2002}, from \eqref{eqn:proof_thoerem1}, we obtain that the mean-field model in \eqref{eqn:mean-field} is globally asymptotically stable.
Therefore, based on \cite[Theorem 3.2]{10.1145/3084454}, we can show the convergences properties of the mean-field model in \eqref{eqn:rates_of_convergence}. We complete the proof of Theorem~\ref{theorem:mean-field}.

\section{Proof of Theorem~\ref{theorem:MFE}} \label{app:MFE}
We prove Theorem~\ref{theorem:MFE} case by case.
For Case 1), if its condition is satisfied, then we must have $1-\frac{\gamma\lambda}{\lambda+\mu} >0$.
Under any given $w<\infty$, by \eqref{eqn:xs}, we have
\begin{align}\label{eqn:xs_inequality}
&x_S^* = \frac{\lambda w (1-\gamma x_S^*)}{(\lambda+\mu)w (1-\gamma x_S^*) + \lambda\mu} < \frac{\lambda w (1-\gamma x_S^*)}{(\lambda+\mu)w (1-\gamma x_S^*)}\nonumber\\
  &\hspace{46mm} =\frac{\lambda}{\lambda + \mu}.
\end{align}
Then, we have
\begin{align}
\frac{C_s}{1-\gamma x_S^*} + \frac{C_t}{\mu}  < \frac{C_s}{1-\frac{\lambda}{\lambda + \mu}} + \frac{C_t}{\mu}   \leq  \left(\frac{1}{\lambda} + \frac{1}{\mu}\right)\hat{C}.
\end{align}
By Lemma~\ref{lemma:optimal_w_for_given_theta},  we obtain that the optimal waiting rate $w^*=\infty$.
Moreover, given that $w^*=\infty$, by \eqref{eqn:xs}, we have $x_S^* = \frac{\lambda}{\lambda+\mu}$. Therefore, we can see that $w^*=\infty$ is the unique MFE.

For Case 2), suppose $w^*<\infty$ is an MFE equilibrium. By Lemma~\ref{lemma:optimal_w_for_given_theta}, we have
\begin{align}
w^* = \frac{ \hat{C}/(1-\theta^*)}{C_s/(1-\theta^*) + C_t/\mu  - \left(1/\lambda + 1/\mu\right)\hat{C}},\label{eqn:proof-w-mfe}
\end{align}
where $\frac{C_s}{1-\theta^*} + \frac{C_t}{\mu}  > (\frac{1}{\lambda} + \frac{1}{\mu})\hat{C}$. Then,  by \eqref{eqn:xs} and $\theta^*=\gamma x_S^*$, we can obtain
\begin{align}
&x_S^* = \frac{\lambda w^* (1-\gamma x_S^*)}{(\lambda+\mu)w^* (1-\gamma x_S^*) + \lambda\mu}\nonumber\\
 &= \frac{\lambda \hat{C}}{(\lambda + \mu)\hat{C} + \lambda\mu(C_s/(1-\gamma x_S^*) + C_t/\mu  - \left(1/\lambda + 1/\mu\right)\hat{C})}\nonumber\\
&=\frac{(1-\gamma x_S^*)\hat{C}}{\mu C_s + (1-\gamma x_S^*) C_t},
\end{align}
which can be further transformed into the following quadratic equation:
\begin{align} \label{eqn:proof-xs-solution}
\gamma C_t (x_S^*)^2 - (\gamma \hat{C}+ \mu C_s + C_t) x_S^* + \hat{C}= 0.
\end{align}
Then, similar to the proof of Theorem~\ref{theorem:mean-field}, we can prove that $(\gamma \hat{C}+ \mu C_s + C_t)^2 -4\gamma C_t\hat{C} > (\gamma\hat{C} + \mu C_s + C_t  - 2 C_t)^2 $ holds, by showing its equivalence to $\mu C_s > 0$.
Thus, the optimal solution to \eqref{eqn:proof-xs-solution} is unique and given by
\begin{align}\label{eqn:solution-xs}
x_S^* =\frac{\gamma\hat{C} + \mu C_s + C_t - \sqrt{(\gamma\hat{C} + \mu C_s + C_t)^2-4\gamma C_t\hat{C}} }{2\gamma C_t} < \frac{1}{\gamma}.
\end{align}
Then, we can easily see that $ w^* = \textcolor{black}{\mathcal{L}_w (\mathcal{L}_{MF}(w^*))}$ and thus, $w^*$ in \eqref{eqn:case2-MNE} is a unique MFE.

\textcolor{black}
{For Case 3), when the condition $\frac{C_s}{1-\theta^*} + \frac{C_t}{\mu}  \leq (\frac{1}{\lambda} + \frac{1}{\mu})\hat{C}$ holds after learning the fraction of busy channels is $\theta^* = \gamma x_S^*$, by Lemma~\ref{lemma:optimal_w_for_given_theta}, all devices will choose the waiting rates $w=\mathcal{L}_w(\bs{x}^*)=\infty$. Then, we can obtain the corresponding fraction of devices in state (S) in the mean-field limit as $\tilde{x}_S = \frac{\lambda}{\lambda + \mu}$.
After learning the fraction of busy channels is $\tilde{\theta} = \gamma\tilde{x}_S = \frac{\gamma \lambda}{\lambda + \mu}$, under the condition $(\frac{1}{\lambda} + \frac{1}{\mu})\hat{C} <\frac{C_s}{\max\{0,1-\frac{\gamma\lambda}{\lambda+\mu}\}} + \frac{C_t}{\mu}$, by Lemma~\ref{lemma:optimal_w_for_given_theta}, we can see that
all devices will change their waiting rates to
\begin{align}
\tilde{w} = \frac{\hat{C}}{C_s + (C_t/\mu  - \left(1/\lambda + 1/\mu\right)\hat{C})\max\{0,1-\frac{\gamma\lambda}{\lambda+\mu}\}}.
\end{align}
Next, we will show that for a given $\tilde{w}$, after learning the corresponding fraction of busy channels, all devices will switch their waiting rates to $w=\infty$.}

\textcolor{black}
{
According to the condition for Case 3), we have $1-\gamma x_S^* > \max\{0,1-\frac{\gamma\lambda}{\lambda+\mu}\}$.
Then, it can be verified that $\tilde{w}<w^*$, where $w^*$ is given by
\begin{align}
w^* = \frac{ \hat{C}/(1-\theta^*)}{C_s/(1-\theta^*) + C_t/\mu  - \left(1/\lambda + 1/\mu\right)\hat{C}}.
\end{align}
Then, we show that $x_S$ in the mean-field limit increases with $w$. According to Theorem~\ref{theorem:mean-field}, we know that
\begin{align}
x_S =\frac{\lambda w (1-\gamma x_S)}{(\lambda+\mu)w (1-\gamma x_S) + \lambda\mu}.
\end{align}
By calculate the derivative of $x_S$ with respect to $w$, we have:
\begin{align}
&\frac{\partial x_S}{\partial w} \nonumber\\
&= \frac{1/\mu}{(1/\lambda+1/(w^2(1-\gamma x_S))+1/\mu)^2}\left(1- \gamma x_S -\gamma w \frac{\partial x_S}{\partial w}\right),\label{eqn:rhs-xs-w}
\end{align}
which implies $\frac{\partial x_S}{\partial w}>0$.
Let $\hat{x}_S$ be the fraction of devices in state (S) in the mean-field limit when the waiting rate is $\tilde{w}$.
Then, we have $\hat{x}_S<x_S^*$, which implies $\frac{C_s}{1-\gamma\hat{x}_S} + \frac{C_t}{\mu} < \frac{C_s}{1-\theta^*} + \frac{C_t}{\mu}  \leq (\frac{1}{\lambda} + \frac{1}{\mu})\hat{C}$.
Then, after learning the fraction of busy channels is $\gamma\hat{x}_S$, by Lemma~\ref{lemma:optimal_w_for_given_theta}, we know that all devices will choose the waiting rate $w=\infty$.
Therefore, we can see that no MFE exist in Case 3), i.e., there does not exist a strategy $w^*$ such that $w^* = \mathcal{L}_w(\mathcal{L}_{MF}(w^*))$, and each device switches its waiting rate between $w=\infty$ and $\tilde{w}$ in \eqref{eqn:case3_w}. We complete the proof of Theorem~\ref{theorem:MFE}.
}

\section{Proof of Proposition~\ref{prop:convergence}} \label{app:convergence}
From the proof of Case 1) of Theorem~\ref{theorem:MFE},  the convergence to the MFE of Case 1) is straightforward.
Thus, we only need to prove the convergence for the MFE of Case 2).
By \eqref{eqn:xs}, for a given $w$, we can see that $\theta=\gamma x_S$ satisfies $\theta = \frac{\gamma\lambda w (1-\theta)}{(\lambda+\mu)w (1-\theta) + \lambda\mu}$.
After tedious calculations, we obtain the derivative of $\theta$ as:
\begin{align}
\frac{\partial \theta}{\partial w} = \frac{\gamma \lambda^2\mu}{((\lambda+\mu)w (1-\theta) + \lambda\mu)^2}\left(1-\theta -w \frac{\partial \theta}{\partial w}\right).
\end{align}
Thus, we have
\begin{align}\label{eqn:theta_w}
\frac{\partial \theta}{\partial w} = \frac{\gamma \lambda^2\mu (1-\theta)}{((\lambda+\mu)w (1-\theta) + \lambda\mu)^2+\gamma \lambda^2\mu w}.
\end{align}
For Case 2), by \eqref{eqn:case2-MNE}, we have
\begin{align}\label{eqn:w_theta}
\frac{\partial w}{\partial \theta} = \frac{\hat{C}(C_t/\mu  - \left(1/\lambda + 1/\mu\right)\hat{C})}{C_s+ (C_t/\mu  - \left(1/\lambda + 1/\mu\right)\hat{C})(1-\theta)}.
\end{align}
It can be easily seen that
\begin{align}
&\left|\frac{\partial w}{\partial \theta}\right|< \frac{\gamma \lambda^2\mu (1-\theta)}{(\lambda\mu)^2} \leq \frac{\gamma}{\mu},\\
&\left|\frac{\partial \theta}{\partial w}\right|< \frac{\hat{C}}{B^2}\left|\frac{C_t}{\mu} - \left(\frac{1}{\lambda} + \frac{1}{\mu}\right)\hat{C}\right|,
\end{align}
 where $B=\min\{C_s,C_s + \frac{C_t}{\mu}  - (\frac{1}{\lambda} + \frac{1}{\mu})\hat{C}\}$.
Therefore, for Case 2), under the condition in \eqref{eqn:convergence-case2}, we have $\left|\frac{\partial }{\partial w}\textcolor{black}{\mathcal{L}_w (\mathcal{L}_{MF}(w))}\right| = \left|\frac{\partial w}{\partial \theta}\right|\left|\frac{\partial \theta}{\partial w}\right|<1$,
which implies the convergence to the MFE in Case 2), according to the contraction mapping theorem. We complete the proof of Proposition~\ref{prop:convergence}.

\bibliographystyle{IEEEtran}
\bibliography{IEEEabrv,ref}

\end{document}